\documentclass[useAMS,usenatbib]{mn2e}
\usepackage{epsfig}

\def\etal{et~al.}
\def\spose#1{\hbox to 0pt{#1\hss}}
\def\lta{\mathrel{\spose{\lower 3pt\hbox{$\mathchar"218$}}
     \raise 2.0pt\hbox{$\mathchar"13C$}}}
\def\gta{\mathrel{\spose{\lower 3pt\hbox{$\mathchar"218$}}
     \raise 2.0pt\hbox{$\mathchar"13E$}}}

\title[Radio--loud AGN in brightest cluster galaxies]{On the prevalence of
radio--loud AGN in brightest cluster galaxies: implications for AGN
heating of cooling flows}

\author[P.~N.~Best \etal]{P.~N.~Best,$^1$\thanks{Email: pnb@roe.ac.uk},
A.~von der Linden$^2$, G.~Kauffmann$^2$, T.~M.~Heckman$^3$, C.~R.~Kaiser$^4$
\\
$^1$ SUPA\thanks{Scottish Universities Physics Alliance}, Institute for
Astronomy, Royal Observatory Edinburgh, Blackford Hill, Edinburgh EH9 3HJ \\
$^2$ Max-Planck-Institut f{\"u}r Astrophysik, Karl-Schwarzschild-Str. 1,
D-85748 Garching, Germany\\
$^3$ Department of Physics \& Astronomy, The Johns Hopkins University,
Baltimore, MD 21218, USA\\
$^4$ School of Physics \& Astronomy, University of Southampton, Southampton
SO17 1BJ\vspace*{-0.4cm}
}

\begin{document}

\pagerange{\pageref{firstpage}--\pageref{lastpage}}
\pubyear{2006}

\label{firstpage}

\maketitle

\begin{abstract}
\noindent The prevalence of radio-loud AGN activity in present-day massive
halos is determined using a sample of 625 nearby groups and clusters
selected from the Sloan Digital Sky Survey. Brightest group and cluster
galaxies (BCGs) are more likely to host a radio--loud AGN than other
galaxies of the same stellar mass (by below a factor of two at a stellar
mass of $\sim 5 \times 10^{11}M_{\odot}$, but rising to over an order of
magnitude below $10^{11}M_{\odot}$). The distribution of radio
luminosities for BCGs does not depend on mass, however, and is similar to
that of field galaxies of the same stellar mass. Neither the radio--loud
fraction nor the radio luminosity distribution of BCGs depends strongly on
the velocity dispersion of the host cluster. The radio-AGN fraction is
also studied as a function of distance from the cluster centre. Only
within 0.2 $r_{\rm 200}$ do cluster galaxies exhibit an enhanced
likelihood of radio--loud AGN activity, which approaches that of the
BCGs. In contrast to the radio properties, the fraction of galaxies with
optical emission--line AGN activity is suppressed within $r_{\rm 200}$ in
groups and clusters, decreasing monotonically towards the cluster centre.

It is argued that the radio--loud AGN properties of both BCGs and non-BCGs
can naturally be explained if this activity is fuelled by cooling from hot
gas surrounding the galaxy. Using observational estimates of the
mechanical output of the radio jets, the time--averaged energy output
associated with recurrent radio source activity is estimated for all
group\,/\,cluster galaxies. Within the cooling radius of the cluster, the
radio--mode heating associated with the BCG dominates over that of all
other galaxies combined. The scaling between total radio--AGN energy
output and cluster velocity dispersion is observed to be considerably
shallower than the $\sim \sigma_v^4$ scaling of the radiative cooling
rate. Thus, unless either the mechanical--to--radio luminosity ratio or
the efficiency of converting AGN mechanical energy into heating increases
by 2--3 orders of magnitude between groups and rich clusters, radio--mode
heating will not balance radiative cooling in systems of all masses. In
groups, radio--AGN heating probably overcompensates the radiative cooling
losses, and this may account for the observed entropy floor in these
systems. In the most massive clusters, an additional heating process (most
likely thermal conduction) may be required to supplement the AGN heating.
\end{abstract}


\section{Introduction}

Brightest cluster galaxies (hereafter BCGs) occupy a unique position at
the centre of the gravitational potential well of clusters of
galaxies. They are amongst the most luminous galaxies in the Universe, up
to ten times brighter than typical elliptical galaxies, with large
characteristic radii \citep[tens of kpc; e.g.][]{sch86} and, in some cases
\citep[the `cD' galaxies; e.g.][]{oem76}, diffuse envelopes which can
extend hundreds of kpc. The properties of BCGs are found to be correlated
with those of their surrounding clusters
\citep[e.g.][]{edg91,lin04,bro05}, and the formation history of BCG and cD
galaxies is believed to be distinct from that of other elliptical
galaxies, due to their special location \citep[e.g.][and references
therein]{ber01,del06}. They are also offset from the standard scaling
relations of early--type galaxies \citep[e.g.][]{lin06}.

Brightest cluster galaxies have long been recognised to show different
radio properties to other cluster galaxies, being much more likely to be
radio--luminous than other non--dominant cluster ellipticals \citep[e.g.\
][]{bur81,val83,bur90}. This radio--loud AGN activity in BCGs has been
proposed as a potential solution to the cooling--flow problem. Gas in the
central regions of clusters of galaxies often has radiative cooling
timescales very much shorter than the Hubble time and, in the absence of a
heating source, a cooling flow would be expected to develop, whereby the
temperature in the central regions of the cluster drops and gas flows
inwards at rates of up to $\sim 1000 M_{\odot}$\,yr$^{-1}$ \citep[see][for
a review]{fab94}. However, recent XMM-Newton and Chandra observations of
cooling flow clusters have shown that the temperature of cluster cores
does not fall below $\sim 30$\% of that at large radii, and that the
amount of cooling gas is only about 10\% of that predicted for a classical
cooling flow \citep[e.g.][]{pet01,dav01,tam01,kaa01}. This implies that
some heating source must balance the radiative cooling losses, preventing
the gas from cooling further.

Heating by radio sources associated with the BCGs has gained popularity in
recent years, as X--ray observations have revealed bubbles and cavities in
the hot intracluster medium of some clusters, coincident with the lobes of
the radio sources \citep[e.g.][]{boh93,car94b,mcn00,fab00b,bla01}.  These
are regions where relativistic radio plasma has displaced the intracluster
gas, creating a low--density bubble of material in approximate pressure
balance with the surrounding medium, which then rises buoyantly and
expands. Hydrodynamic simulations have been able to reproduce bubbles with
properties similar to those observed \citep[e.g.\
][]{chu01,qui01,bru02,bru03b} and have also shown that, provided that
radio source activity is recurrent, the total energy provided by AGN
activity can be sufficient to balance the cooling radiation losses through
repeated production of jets, buoyant bubbles and associated shocks
\citep[e.g.][]{dal04,bru05,nus06}. Other authors \citep[e.g.\
][]{omm04,bri06} have argued that momentum--driven jets are an alternative
means of distributing AGN energy throughout the intracluster medium.

Very deep Chandra observations of the Perseus and Virgo clusters
\citep{fab03,fab06,for05,for06} have revealed the presence of
approximately spherical weak shock waves in these clusters, extending out
to several tens of kpc radius. \citet{fab03} argued that these `ripples'
are excited by the expanding radio bubbles, and that dissipation of their
energy can provide a quasi--continuous heating of the X--ray gas
\citep[see also][]{rus04}. \citet{fuj05} and \citet{mat06} argued,
however, that if all of the wave dissipation occurs at the shock front
then too much of the heat is deposited in the inner regions of the
cluster, and not enough out at the cooling radius, leading to temperature
profiles at variance with observations. \citet{nus06} suggested that a
natural solution to this problem is to consider the heating effects of AGN
activity from all cluster galaxies, not only BCGs. More recently,
\citet{fab06} showed that the shocks occurring in the Perseus cluster are
isothermal, meaning that thermal conduction must be important in mediating
the shock. They argue that this prevents the accumulation of hot shocked
gas in the inner regions, and that the energy of the waves can be
dissipated at larger radii by viscous effects.

One critical question for all of these models is whether the heating
generated by an episode of radio--loud AGN activity is sufficient, during
its lifetime, to offset the radiative cooling losses of the cluster.
Viewed in terms of the radio bubbles (which are widely considered to be
the driving force behind the shock and sound waves), for some clusters the
(pV) energy contained within the evacuated bubbles has been shown to be
sufficient to balance the cooling losses, at least for a short period of
time \citep[a few $\times 10^7$\,yr; e.g.\
][]{fab03,bir04,dun05}. However, \citet{bir04} showed that this is not
true for about half of the clusters they studied \citep[see
also][]{raf06}.  \citet{nus06} and others have argued that the mechanical
energy injected into the cluster by weak shocks may be up to an order of
magnitude larger than pV, which would ease this problem. An alternative
possibility is that the heat balance occurs in a quasi--static manner,
with recurrent low luminosity radio activity being punctuated by
occasional major eruptions supplying much more energy \citep{kai03}.

A second important question is what the duty cycle of this AGN activity
is. The duty cycle determines the rate of production of radio bubbles, or
equivalently, the timescale between the sound wave `ripples', and hence is
required to calculate the time--averaged heating rate associated with AGN
activity. \citet{bur90} showed that as many as 70\% of cD galaxies at the
centre of cooling flow clusters are radio--loud, but this result was based
on a sample of only 14 such systems, and not all BCGs are cD galaxies.

Recently, \citet{bes05b} used data from the Sloan Digital Sky Survey
\citep[SDSS;][]{yor00, sto02} to investigate the origin of radio--loud AGN
activity, and found that the dominant factor which determined whether or
not a galaxy is radio--loud is its mass: the fraction of galaxies which
were found to host radio--loud AGN scaled as $M_*^{2.5}$ or $M_{\rm
BH}^{1.6}$, where $M_*$ and $M_{\rm BH}$ are the stellar and black hole
masses of the galaxy, respectively. The distribution of radio luminosities
was found to be the same regardless of the galaxy mass. Combining these
results with a conversion between radio luminosity and the mechanical
energy supplied by a jet to its surroundings \citep[as estimated by
][]{bir04}, \citet{bes06a} estimated the time--averaged mechanical
luminosity output associated with radio source activity for each galaxy,
as a function of its mass. They found that, for elliptical galaxies of all
masses, the time averaged radio heating almost exactly balanced the
radiative cooling losses from the hot haloes of the ellipticals. They
argued that the radio AGN feedback may therefore play a critical role in
galaxy formation, preventing further cooling of gas onto elliptical
galaxies and thus explaining why these galaxies are old and red \citep[see
also][]{ben03,sca05,cro06,bow06,cat06}.

\citet{bes06a} also concluded that, unless brightest cluster galaxies
showed a different mode of radio activity to that of ordinary elliptical
galaxies, then they could not provide sufficient heating to balance the
cooling in clusters of galaxies. As discussed above, BCGs {\it are} found
to be radio--luminous much more frequently than other cluster ellipticals,
but the strong mass dependence of the radio--loud AGN fraction found by
\citet{bes05b} means that this may solely be due to their very much higher
masses. The goal of the current paper is therefore to investigate this
issue, by determining both the radio--loud fraction, and the distribution
of radio luminosities, for a large sample of brightest group and cluster
galaxies with a wide range of masses and host group\,/\,cluster
properties. This will then ascertain the importance of radio--AGN heating
in clusters. A secondary goal of the paper is to investigate the role that
non--BCG cluster galaxies may have in heating the intracluster medium.

The layout of the paper is as follows. Section~\ref{samples} describes the
various data samples that are used for this analysis.  An analysis of the
radio properties of brightest group and cluster galaxies is carried out in
Section~\ref{radbcg}, while Section~\ref{radclus} investigates the radio
properties of the other group and cluster galaxies. These are compared
against signatures of optical AGN activity in Section~\ref{optagn}. The
implications of these results for the origin and fuelling of low
luminosity radio source activity are discussed in Section~\ref{implics}.
In Section~\ref{cooling} the radio--AGN heating rates are estimated in
clusters and groups across a wide range of masses, and compared these with
the radiative cooling rates. The implications of these results for AGN
heating of cooling flows are discussed in Section~\ref{impli2}, and
conclusions are drawn in Section~\ref{concs}. Throughout the paper, the
cosmological parameters are assumed to have values of $\Omega_m = 0.3$,
$\Omega_{\Lambda} = 0.7$, and $H_0 = 70$\,km\,s$^{-1}$Mpc$^{-1}$.

\section{Sample selection}
\label{samples}

\subsection{Definition of the galaxy samples}
\label{clussamp}

The Sloan Digital Sky Survey \citep{yor00, sto02} is a five-band
photometric and spectroscopic survey which will ultimately cover about a
quarter of the extragalactic sky; the fourth data release of this survey
\citep[DR4;][]{ade06} includes spectroscopy of over half a million
objects. The parent sample for the current study is drawn from the `main
galaxy sample' \citep{str02} of the SDSS DR4, comprising those galaxies
with magnitudes in the range $14.5 < r < 17.77$. The spectra of these
galaxies have been used to derive a large number of physical properties,
with catalogues of measured and derived parameters being publically
available (see http://www.mpa-garching.mpg.de/SDSS/).

\citet{mil05} have used the SDSS data to identify clusters of galaxies as
overdensities in a seven-dimensional space of position and colour. This
`C4' cluster catalogue is currently available for DR3, and identifies 1106
clusters with redshifts $0.02 < z < 0.16$.  Miller \etal\ also identify
two candidate brightest cluster galaxies for each of their clusters: the
`mean galaxy' is that closest to the peak of the galaxy density field, and
the `brightest galaxy' is the brightest spectroscopically confirmed
cluster member with a projected position within 500$h^{-1}$kpc (where $h$
= $H_0 / 100$\,km\,s$^{-1}$Mpc$^{-1}$) of that density peak and a redshift
within four times the velocity dispersion of the mean cluster redshift. In
many cases, however, neither of these turns out to be the true BCG
\citep[see][]{lin06}. For example, in about 30\% of dense clusters, the
BCG is not included in the spectroscopic data, due to the problems of
fibre collisions, and so is missed by the C4 catalogue.

In a companion paper, \citet{lin06} have re-analysed the C4 cluster
catalogue in order to provide a robust sample of clusters with
well--defined BCGs. The reader is referred to that paper for a detailed
discussion of how the cluster sample and BCGs were defined. In brief, the
sample was first restricted to the 833 C4 clusters with redshifts $z <
0.1$. The brightest cluster galaxies were then identified using all
available information (magnitude, colour, morphology, redshift if in the
spectroscopic sample), together with visual inspection of colour images of
the clusters (see von der Linden \etal\ for details). During this process,
clusters which were identified to be substructures of larger clusters were
removed from the sample, as were those for which an iterative algorithm to
determine the cluster redshift, velocity dispersion ($\sigma_{\rm v}$) and
virial radius ($r_{\rm 200}$) either did not converge or retained three or
fewer galaxies within 3$\sigma_{\rm v}$ and 1$r_{\rm 200}$ of the cluster
centre. This resulted in a final sample of 625 systems, for which the
size, velocity dispersion, BCG and cluster membership were all well
determined.  It turns out that some of these systems have velocity
dispersions which indicate that they are galaxy groups rather than
clusters; the terms `cluster' and `brightest cluster galaxy' are used
loosely in this paper, to include both clusters and groups. 

A further critical re--analysis carried out by \citet{lin06} has been to
correct the SDSS photometry for these BCGs. The SDSS photometry
systematically underestimates the luminosities of nearby large galaxies,
particularly in cluster environments, because it overestimates the level
of sky background \citep{ber06,lau06,lin06}. This then leads to
underestimates of the galaxy mass, which would impact upon the current
studies. von der Linden \etal\ derived a method to correct the photometry
for this sky over-subtraction; their method uses both the `local' and
`global' sky estimates provided by the SDSS pipeline, combined according
to a weighted ratio of the luminosity of the galaxy to that of its
neighbouring galaxies. They show that this method is successful in
reproducing magnitudes which match with independent measurements (see
their paper for more details).

In the current paper, the `brightest cluster galaxy' sample generally
corresponds to the 625 BCGs defined from this sample of clusters. In 141
of these clusters, however, the BCG is not contained within the
spectroscopic catalogue; therefore, where comparison is being made with
properties derived from spectroscopic observations (e.g. emission line
strengths), the sample is restricted to only those 484 BCGs with
spectroscopic data. Since the BCGs with only photometric data have
generally been excluded from the spectroscopic sample due to random
effects (especially fibre collisions) it is not expected that this will
produce any significant bias in the sample.

The results for the BCGs are compared with those of `all galaxies'. The
`all galaxy' sample corresponds to those galaxies within the main galaxy
sample of the SDSS DR4 release, with redshifts $0.02 < z < 0.1$; these
redshift cuts match those applied to the cluster catalogue, and von der
Linden \etal\ have also applied their photometric correction method to
these galaxies in an equivalent way. The all galaxy sample is discussed by
von der Linden \etal: it includes a mix of galaxy properties, in a wide
range of environments, but in the luminosity range where the sample
overlaps with that of the BCGs it is dominated by early--type galaxies. 

A `non-BCG cluster galaxy' sample is also constructed. This consists of
those galaxies within the `all galaxy' sample which are within 3$r_{200}$
and 3$\sigma_{\rm v}$ of one of the 625 clusters identified by von der
Linden \etal, but which are not identified as a BCG.

For all galaxies, the rest-frame luminosities were calculated from the
re-calibrated photometry using the {\sc kcorrect} algorithm \citep{bla03},
which determines the best composite fit to the observed galaxy fluxes of a
large number of template stellar population spectra, including three
alternative models for dust extinction.  Following \citet{bla07}, this
same algorithm was used to derive stellar mass estimates: as shown by
those authors, the stellar masses derived agree to typically within 0.1
dex of those derived by other methods \citep[e.g. those of][using the
z--band magnitude and a mass--to--light ratio determined by the strengths
of the 4000\AA\ break and the $H\delta$ absorption in the galaxy
spectrum]{kau03a}. Other physical properties of the galaxies, such as
their emission line strengths, were taken from the MPA pipeline catalogues
available at {\rm http://www.mpa-garching.mpg.de/SDSS/}.

\subsection{Identification of radio--loud AGN}
\label{radsamp}

\citet{bes05a} identified the radio--emitting galaxies within the main
spectroscopic sample of the SDSS DR2, by cross--comparing these galaxies
with the National Radio Astronomy Observatory (NRAO) Very Large Array
(VLA) Sky Survey \citep[NVSS;][]{con98} and the Faint Images of the Radio
Sky at Twenty centimetres (FIRST) survey \citep{bec95}. The use of a
combination of these two radio surveys allowed a radio sample to be
derived which was both reasonably complete ($\approx 95$\%) and highly
reliable ($\approx 99$\%). They then used the optical properties of the
galaxies to separate the radio--loud AGN from the radio--detected
star--forming galaxies. This work has now been extended to include the DR4
data (Best et~al. in prep), and these results were used to determine which
galaxies from the SDSS spectroscopic samples are radio--loud AGN.

\begin{figure}
\centerline{
\psfig{file=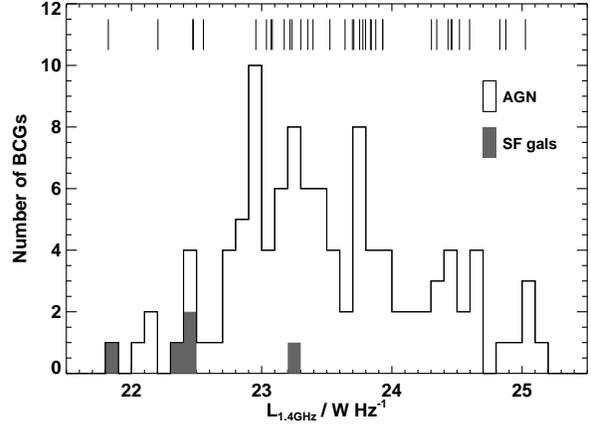,angle=0,width=8.6cm,clip=}
}
\caption{\label{lhistbcg} The histogram shows the distribution of radio
luminosities of the radio--detected BCGs with SDSS spectroscopic data,
split by their classification as either radio--loud AGN (unshaded) or
radio--detected star--forming galaxies (shaded). The vertical lines at the
top of the plot show the radio luminosities of the BCGs with only
photometric data.}
\end{figure}

For the 141 brightest cluster galaxies with only photometric data, the
same cross--comparison of the galaxy locations with the NVSS and FIRST
radio surveys was repeated in order to determine which had associated
radio emission. However, the lack of spectroscopic data for these galaxies
meant that the previous method of distinguishing radio--loud AGN from
star--forming galaxies could not be used. Instead, for the 37 photometric
BCGs with radio detections, the classification was based solely upon the
radio luminosity. Figure~\ref{lhistbcg} shows the distribution of radio
luminosities for the BCGs with spectroscopic data, split by their
classification as either radio--loud AGN or radio--detected star--forming
galaxies. Above $L_{\rm 1.4GHz} = 10^{22.5}$W\,Hz$^{-1}$ all but one radio
detected BCGs are AGN, whereas below that luminosity nearly half are star
forming galaxies. The vertical lines at the top of the plot indicate the
radio luminosities of the BCGs without spectroscopic data. The 32
photometric BCGs with $L_{\rm 1.4GHz} > 10^{22.9}$W\,Hz$^{-1}$ are almost
certainly radio--loud AGN, and henceforth are classified as such. The
other 5 all have radio luminosities below $10^{22.6}$W\,Hz$^{-1}$ and are
likely to include a mixture; their classification is less critical because
most analyses are limited to only the BCGs brighter than
$10^{23}$W\,Hz$^{-1}$, but conservatively it is assumed that these are all
star forming galaxies. It should be emphasised that the number of BCGs for
which the origin of the radio emission is ambiguous is far too small for
any misclassification to influence the results of the paper.

\section{The radio activity of brightest cluster galaxies}
\label{radbcg}

Using the radio sources defined from the SDSS DR2 catalogue,
\citet{bes05b} showed that the probability of a galaxy to be a radio--loud
AGN was dependent primarily upon its mass, determined either as the
stellar mass ($f_{\rm radio-loud} \propto M_*^{2.5}$), or the black hole
mass ($f_{\rm radio-loud} \propto M_{\rm BH}^{1.6}$). As discussed by
\citet{bes05b}, the difference between the slopes of these two relations
arises mainly because of the increasing fraction of disk--dominated
galaxies, with small black holes, at stellar masses $M_* \lta
10^{11}M_{\odot}$. These host fewer radio--loud AGN, and thus decrease the
radio--loud fraction at low stellar masses, leading to a steeper
dependence on stellar mass.

\begin{figure}
\centerline{
\psfig{file=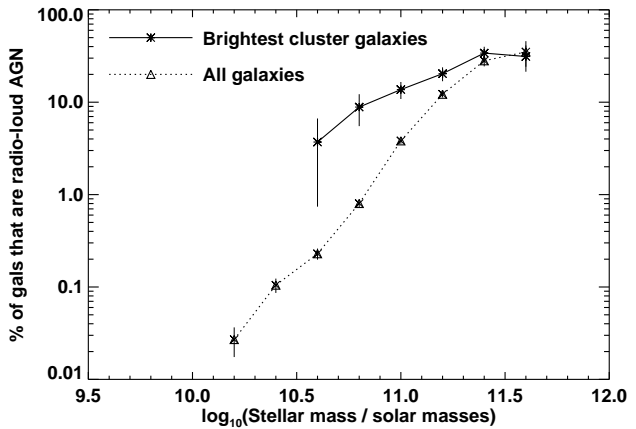,angle=0,width=8.6cm,clip=}
}
\caption{\label{bcgradfrac} The percentage of galaxies that are
radio--loud AGN, as a function of stellar mass, for `all galaxies' and for
`brightest cluster galaxies'. Brightest cluster galaxies are more likely
to be radio--loud than other galaxies of the same stellar mass, and the
two relations have different slopes. }
\end{figure}

Figure~\ref{bcgradfrac} shows the percentage of brightest cluster galaxies
that are radio--loud AGN, as a function of stellar mass. For comparison,
the equivalent relation is also shown for the all galaxy sample. The
brightest cluster galaxies are clearly offset from the `all galaxy'
relation, and their relation shows a different slope\footnote{It should be
stressed that the same conclusions are reached if the plots are made as a
function of galaxy luminosity, instead of stellar mass, as demonstrated in
\citet{lin06}. Similarly, if the `all galaxy' sample is restricted to just
those classified as early-type galaxies by using their SDSS morphological
parameters concentration index and surface mass density \citep[cf.][and
discussion therein]{kau03b} then the results are also largely unchanged at
masses above $5 \times 10^{10} M_{\odot}$, where it overlaps with the BCG
sample. This is because the galaxies at high stellar masses are dominated
by old ellipticals, whose mass-to-light ratios do not vary strongly.}: the
fraction of BCGs that is radio--loud increases roughly as $M_*^{1.0}$ and
is an order of magnitude higher than that of all galaxies for stellar
masses below $10^{11} M_{\odot}$, but comparable to that of all galaxies
at stellar masses above $5 \times 10^{11} M_{\odot}$. The increasing
similarity of the two radio-loud fractions with increasing mass is
actually not surprising, since at masses above $10^{11.5}M_{\odot}$ around
a third of the `all galaxy' population are classified as BCGs, and many of
the remainder may be BCGs missed by the current sample. Note that from
these data it is not possible to tell whether the increase in the
radio--loud fraction of BCGs, relative to that of other galaxies, is due
to radio--AGN activity being triggered more frequently in BCGs, or the BCG
radio sources being longer lived (e.g. because the higher pressure
environment of the dense cluster environment slows down the source
expansion).

Black hole mass may be considered to be a more fundamental parameter than
stellar mass when considering nuclear activity, and \citet{bes05b}
estimated this for the SDSS DR2 galaxies using the velocity dispersion
versus black hole mass relation of \citet{tre02}: log$(M_{\rm BH} /
M_{\odot}) = 8.13 + 4.02 \rm{log}(\sigma_*/200$km\,s$^{-1})$.  However,
there is some doubt as to whether this `normal galaxy' relation is also
applicable for BCGs, or whether these follow a different relation
\citep[e.g.][]{lin06}.  In addition, velocity dispersion measurements are
only available for those BCGs with spectroscopic observations. In order to
avoid any potential biases therefore, and also to increase the sample
size, in this paper the analysis is presented only as a function of
stellar mass. As a check, relations were constructed as a function of
black hole mass, assuming the velocity dispersion to black hole mass
conversion to be the same as for normal galaxies; these are in broad
agreement with those presented here for stellar mass.

\begin{figure}
\centerline{
\psfig{file=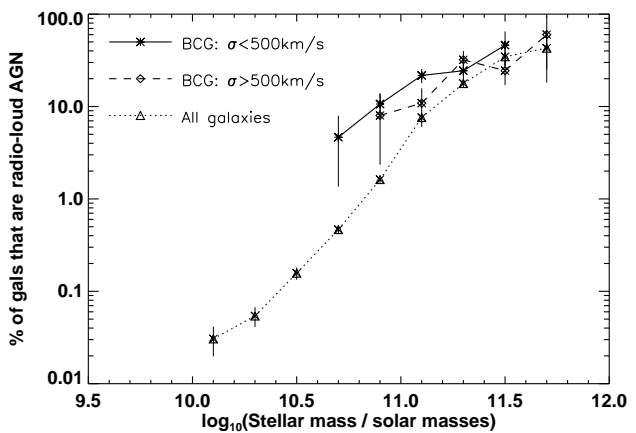,angle=0,width=8.6cm,clip=}
}
\caption{\label{bcgradsigma} The percentage of brightest cluster galaxies
that are radio--loud AGN, as a function of stellar mass, split by the
velocity dispersion of the clusters. At given stellar mass, the
radio--loud fraction of BCGs is indistinguishable between rich and poor
clusters.}
\end{figure}

Figure~\ref{bcgradsigma} compares the radio--loud fraction versus stellar
mass relation for BCGs in rich versus poor clusters. The probability for a
BCG of given stellar mass to host a radio--loud AGN is independent of the
velocity dispersion of the surrounding cluster. Because BCGs in the
richest clusters tend to be more massive than those in poorer clusters,
the fraction of BCGs which host radio--loud AGN still turns out to be
higher in more massive clusters (marginally at least; $21 \pm 2$\% in the
$\sigma_{\rm v} > 500$km\,s$^{-1}$ clusters in this sample, compared to
$17 \pm 2$\% for those with $\sigma_{\rm v} < 500$km\,s$^{-1}$), but it is
the mass of the BCG, rather than the properties of its surroundings, which
controls the likelihood of it being radio--loud.

\begin{figure}
\centerline{
\psfig{file=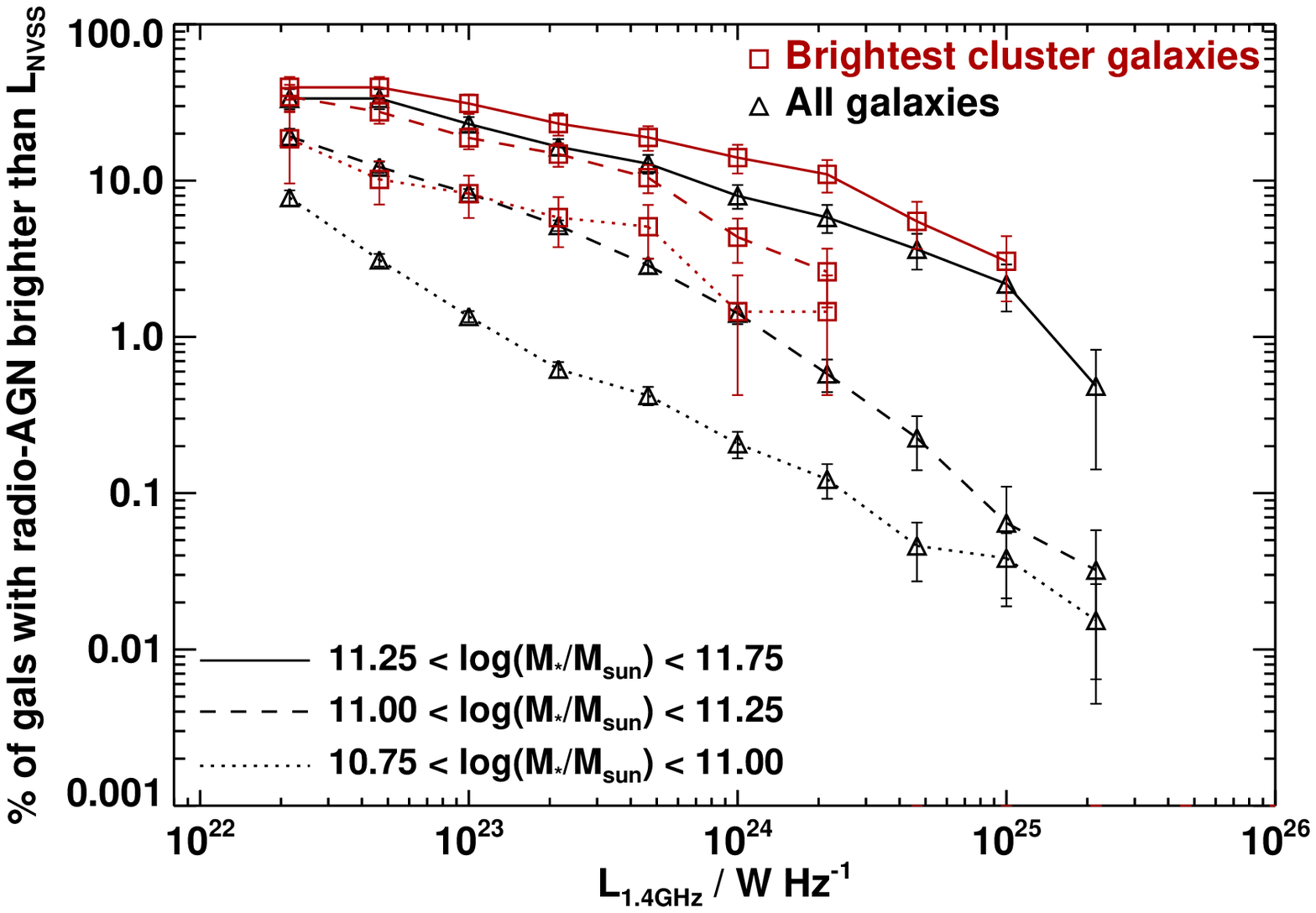,angle=0,width=8.6cm,clip=}
}
\centerline{
\psfig{file=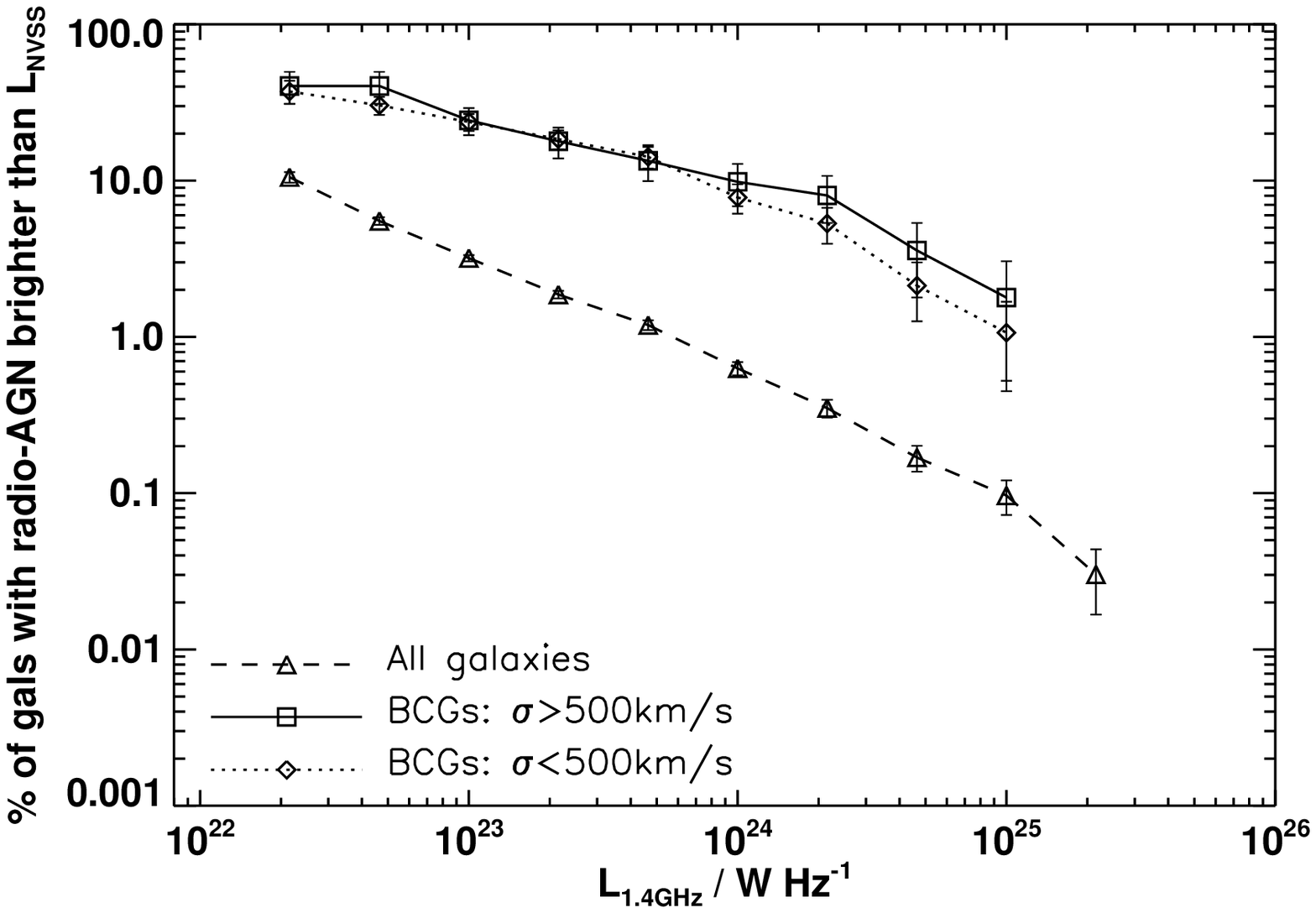,angle=0,width=8.6cm,clip=}
}
\caption{\label{radfuncmass} {\it Top:} The percentage of galaxies which
are radio--loud AGN brighter than a given radio luminosity, for `all
galaxies' (black lines) and the subsample of `brightest cluster galaxies'
(red lines) in three different ranges of stellar mass. There is no
evidence for any significant difference in the shape of the luminosity
functions, either as a function of stellar mass, or between brightest
cluster galaxies and all galaxies: only the normalisation of the relation
changes. {\it Bottom:} The distribution of radio luminosities of BCGs
split by the velocity dispersion of their host clusters, for BCGs with
masses in the range $10^{11} < M_* / M_{\odot} < 5 \times 10^{11}$,
compared to that of normal galaxies.}
\end{figure}

\citet{bes05b} also found that for `all galaxies' the distribution of
radio luminosities of the radio--loud AGN was independent of the mass of
the host galaxy. This result is shown in the top panel of
Figure~\ref{radfuncmass} for three different mass ranges.  Also shown on
the same plot are the distributions of radio luminosities for brightest
cluster galaxies in the same mass ranges. The shape of the radio
luminosity function does not change significantly either as a function of
mass for brightest cluster galaxies, or between the brightest cluster
galaxy and all galaxy samples\footnote{The only variation in shape is a
flattening at low radio luminosities, of both BCGs and all galaxies, when
the radio--loud fraction exceeds $20-30$\%; as \citet{bes05b} argued,
such flattening must occur at some point, since the radio--loud fraction
cannot exceed 100\%.}: only the normalisation of the radio luminosity
function changes. The bottom panel of Figure~\ref{radfuncmass} compares
the distribution of radio luminosities of BCGs in rich and poor clusters:
again, these are broadly in agreement, although there is tentative
evidence that there may be more high luminosity radio sources in richer
clusters. A larger sample will be required to confirm this result.

\section{Radio--loud AGN activity in other cluster galaxies}
\label{radclus}

Many authors have examined the environments of radio--loud AGN
\citep[e.g.][]{pre88,hil91,mil02b,bes04a} and have found that these
sources appear to favour galaxy group and weak cluster
environments. Nevertheless, \citet{led96} demonstrated that the bivariate
radio--optical luminosity function of cluster galaxies is very similar to
that of field galaxies. This indicates that the preference for radio
sources to be located in dense environments may simply be due to the
concentration of massive galaxies within clusters, coupled with the strong
mass dependence of the radio--loud fraction. The SDSS data and the C4
cluster sample allow this issue to be investigated in more detail.

\begin{figure}
\centerline{
\psfig{file=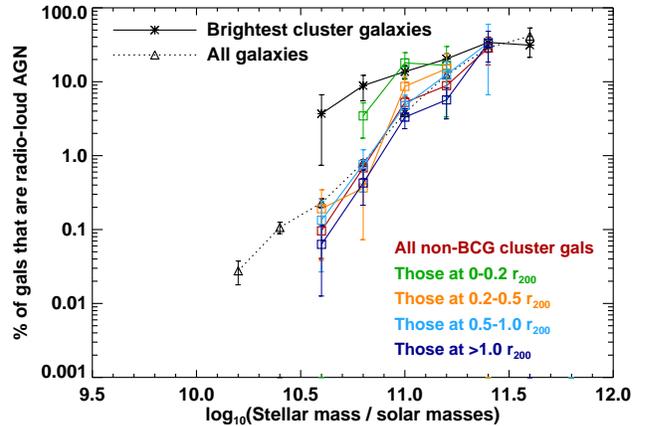,angle=0,width=8.6cm,clip=}
}
\caption{\label{clusradfrac} The percentage of galaxies that are
radio--loud AGN, as a function of stellar mass, for `non--BCG cluster
galaxies' (red line) as compared to the `all galaxy' (dotted black line)
and `brightest cluster galaxy' (solid black line) samples. The other
coloured lines indicate the radio--loud AGN fractions for subsets of the
non--BCG cluster galaxies within different ranges of projected
cluster-centric radii: the green, orange and cyan lines show the relations
for galaxies with redshifts within 2$\sigma_{\rm v}$ of the mean cluster
redshift and with projected radii of 0--0.2$r_{200}$, 0.2--0.5$r_{200}$
and 0.5--1.0$r_{200}$ respectively, while the dark blue line shows the
equivalent relation for cluster galaxies outwith 1.0$r_{200}$.}
\end{figure}

\begin{figure}
\centerline{
\psfig{file=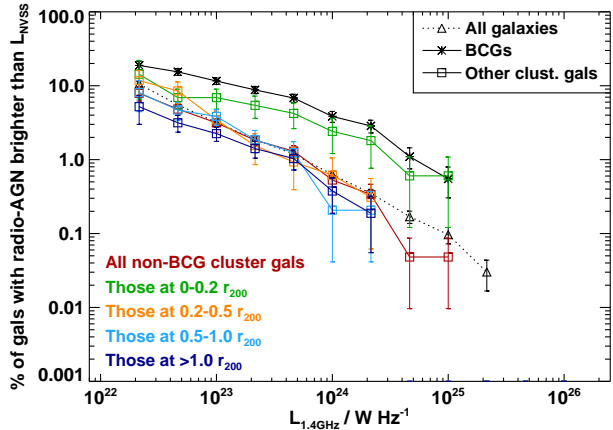,angle=0,width=8.6cm,clip=}
}
\caption{\label{radfuncclus} The percentage of galaxies which are
radio--loud AGN brighter than a given radio luminosity for `non--BCG
cluster galaxies' (red line) as compared to the `all galaxy' (dotted black
line) and `brightest cluster galaxy' (solid black line) samples. These
plots are constructed considering all galaxies in the $10^{11} < M_* /
M_{\odot} < 10^{12}$ mass range, but then scaling the BCG line uniformly
down by a factor of 1.7 to account for the higher median mass of BCGs
within this mass range (the median masses of galaxies in all other
subsamples are statistically indistinguishable). The other coloured lines
indicate the equivalent relations for different subsets of the non--BCG
cluster galaxies with different projected radii, as defined in
Figure~\ref{clusradfrac}. The radio--loud AGN fraction is boosted for
galaxies within 0.2$r_{200}$ of the cluster centre, but retains the same
overall shape.}
\end{figure}

\begin{figure}
\centerline{
\psfig{file=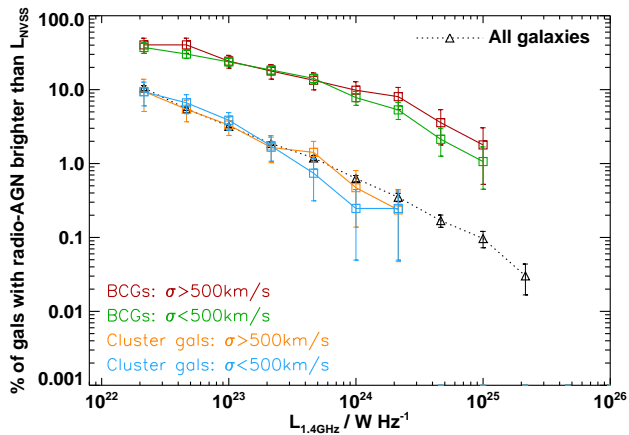,angle=0,width=8.6cm,clip=}
}
\caption{\label{radfuncclussig} The distribution of radio luminosities of
non--BCG cluster galaxies, split by the velocity dispersion of their host
clusters, as compared to all galaxies and to brightest cluster galaxies.
Only galaxies with masses in the range $10^{11} < M_* / M_{\odot} < 5
\times 10^{11}$ are included on the plot.}
\end{figure}

Figure~\ref{clusradfrac} shows the fraction of non--BCG cluster galaxies
(red line) that are radio--loud AGN, as a function of stellar mass, in
comparison to all galaxies and to brightest cluster galaxies. The
radio--loud AGN fraction of cluster galaxies matches that of all galaxies
at all stellar masses (except possibly the lowest mass bin, where it might
be slightly lower).  Figure~\ref{radfuncclus} similarly shows that the
distribution of radio luminosities of the cluster galaxies is the same as
that of non-cluster galaxies. Overall, then, for galaxies of given stellar
mass the radio luminosity functions are the same inside and outside of
clusters, as was found by \citet{led96}. They are also independent of the
velocity dispersion of the host cluster, as shown in
Figure~\ref{radfuncclussig}.

Figures~\ref{clusradfrac} and~\ref{radfuncclus} also show the radio--loud
AGN fraction and radio luminosity distribution for various subsets of
cluster galaxies in different ranges of projected radius from the cluster
centre. It is striking that galaxies with redshifts within 2$\sigma_{\rm
v}$ of the mean cluster redshift and 0.2$r_{200}$ of the cluster centre
have an enhanced probability of being radio--loud AGN, approaching that of
the BCGs, although once again with the same distribution of radio
luminosities. Outside of 0.2$r_{200}$ there is no substantial change in
the radio--loud fraction with radius (there may be a marginal fall with
increasing radius, but this is within the errors, and insignificant
compared to the increase within 0.2$r_{200}$).

One concern is that the increase in the radio--loud fraction of galaxies
near the centre of the cluster may arise as a result of mis-identification
of some BCGs. The evidence suggests, however, that this is not the
case. There are 13 clusters where a radio source is associated with a
non--BCG cluster galaxy within 0.2$r_{200}$. Of these, the identified BCG
is also radio--loud in 6 ($\equiv 46 \pm 19$\%) of the clusters: this
fraction is comparable to (or even above) that of other BCGs of the same
mass ($\approx 10^{11.3} M_{\odot}$), and far above that of non--BCG
cluster galaxies. Further, of the 7 clusters with a radio--loud galaxy
within 0.2$r_{200}$ but a radio--quiet BCG, visual analysis indicates that
the BCG identification is absolutely unambiguous in 5 of the cases, and
should be reliable in the other two. Finally, the distribution of ratios
of radio galaxy mass to BCG mass is indistinguishable between the 13 $r <
0.2r_{200}$ radio galaxies and the radio galaxies at larger radii, in
contrast to what would be expected if these were misidentified BCGs. In
conclusion, therefore, the enhanced radio--loud fraction for galaxies
within $0.2r_{200}$ of the cluster centre appears to be a genuine physical
effect.

\section{Emission--line AGN activity in cluster galaxies}
\label{optagn}

\begin{figure}
\centerline{
\psfig{file=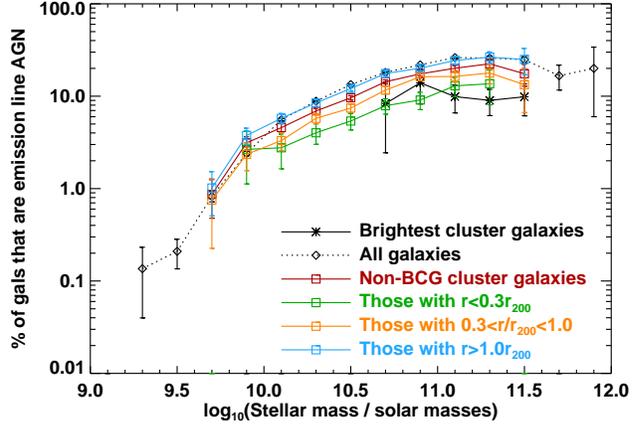,angle=0,width=8.6cm,clip=}
}
\caption{\label{optfracall} The percentage of galaxies that show
emission--line AGN activity (with $L_{\rm [OIII] 5007} \gta 10^{6.5}
L_{\odot}$), as a function of stellar mass, for `all galaxies' (dotted
black line), `brightest cluster galaxies' (solid black line) and `non--BCG
cluster galaxies' (red line). Also shown are equivalent relations for
subsamples of cluster galaxies in different cluster-centric radial
ranges. Emission line AGN activity is suppressed within $r_{200}$ of the
cluster centre, with the strength of the suppression increasing with
decreasing radius.}
\end{figure}

It is interesting to compare the enhanced radio--loud AGN activity of
brightest cluster galaxies and other galaxies in the central $0.2r_{200}$
of the cluster, with their emission line properties. \citet{kau03c}
identified emission--line galaxies in the SDSS spectroscopic sample, and
used the [OIII]~5007 / H$\beta$ versus [NII]~6583 / H$\alpha$ emission
line ratio diagnostic diagram \citep{bal81} to separate out galaxies with
AGN activity from those galaxies where the emission lines are associated
with star formation\footnote{It should be stressed that since the parent
sample for that analysis, like the current one, was the SDSS `main galaxy
sample', the AGN selected are mostly type--II AGN. Powerful type-I quasars
will be missed from the sample.}. The fraction of galaxies hosting
emission line AGN rises shallowly up to $10^{10.5} M_{\odot}$, and is then
flat at higher masses, which is substantially different from the very
steep mass dependence of the radio--loud AGN fraction \citep{bes05b}. This
result is shown in Figure~\ref{optfracall} for the `all galaxy' sample
\citep[for AGN defined in the same way as][and with $L_{\rm [OIII] 5007} \gta
10^{6.5} L_{\odot}$]{kau03c} and is compared to the fraction of galaxies
hosting emission--line AGN for the brightest cluster galaxy and non-BCG
cluster galaxy samples. Also shown are the relations for subsamples of
cluster galaxies at different cluster-centric radii. Note that once again,
making this plot as a function of galaxy luminosity instead of mass leads 
to essentially the same results.

Brightest cluster galaxies are found to be less likely to possess
emission--line AGN activity than `all galaxies' of the same stellar mass,
by a factor of 2--3 at all masses. This suppression of emission--line AGN
activity is also true of other cluster galaxies: cluster galaxies as a
whole are found to be about 20\% less likely to show emission--line AGN
activity than all galaxies, consistent with the results of previous
studies \citep[e.g.][]{mil03a,kau04}.  Splitting the cluster galaxies into
subsamples at different radii demonstrates that the suppression gets
progressively stronger as the cluster centre is approached: outside 
$r_{200}$ the fraction of cluster galaxies displaying emission--line AGN
activity matches that of `all galaxies', but this drops by a factor of at
least 2 for those galaxies within $0.3 r_{200}$. Interestingly the BCGs
show the same likelihood of emission line AGN activity as other galaxies
of the same mass towards the centre of the cluster: unlike for the radio
activity, the special location of BCGs does not appear to lead to
different emission--line AGN activity. 

This result may seem to be at variance with the luminous extended emission
line structures known to exist around some BCGs \citep[e.g.][and
references therein]{hec89,hat06}. Part of this difference may be because
the SDSS fibre apertures are small ($\sim 5$kpc diameter at $z \approx
0.1$), whilst the line emission seen in the above studies can extend for
several tens of kiloparsec. More important, however, is a difference in
the type of BCG studied. \citet{edw07} recently studied the line emission
from BCGs and found that the fraction of the BCG population as a whole
that display emission lines is about 15\%, and is comparable to that of
other massive galaxies near the cluster centres. These results are in full
agreement with those found here, shown in Figure~\ref{optfracall}.
However, these authors also find that if they restrict analysis to just
those BCGs which are within 50\,kpc of the X--ray centre of a cooling flow
cluster, then the fraction which display emission lines rises to $\sim
75$\%.  It is such massive, strong cooling flow clusters that have been
the main focus of the previous emission line studies, and the extended
emission line activity detected is associated with the cooling flow. The
current results and those of Edwards et~al. demonstrate that more typical
BCGs do not have enhanced emission line activity.

\begin{figure}
\centerline{
\psfig{file=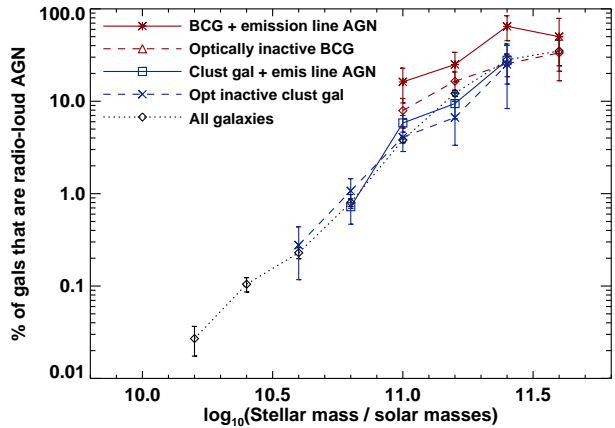,angle=0,width=8.6cm,clip=}
}
\caption{\label{optpassall} The percentage of brightest cluster galaxies
which are also emission--line selected AGN (solid red line) and those
which are optically--inactive (dashed red line), that are classified as
radio--loud AGN, as a function of stellar mass. This is compared to the
fraction of normal galaxies that are radio--loud AGN (black line; this is
independent of whether or not they optical AGN), and with the equivalent
relations for non-BCG cluster galaxies (blue lines).}
\end{figure}

A further interesting test concerns the relationship between radio and
optical emission line activity. \citet{bes05b} found that for `all
galaxies' the probability of a galaxy hosting a low luminosity radio
source was independent of whether or not it was an emission--line AGN.
Figure~\ref{optpassall} investigates this for the brightest cluster
galaxies and the non--BCG cluster galaxies. Unlike the situation for all
galaxies, the radio--loud fraction for brightest cluster galaxies does
depend upon the emission line properties: BCGs which show emission--line
AGN activity are 2--3 times more likely at all masses to host radio--loud
AGN than those which are optically inactive --- although the latter still
show a higher radio--loud fraction than non--BCG galaxies. Other cluster
galaxies behave in the same way as `all galaxies', with the radio--loud
fraction being the same regardless of the optical properties.

\section{Implications for the origin of low luminosity radio-loud AGN}
\label{implics}

The apparent independence of emission line and radio--loud AGN activity,
coupled with the fundamentally different dependencies of their prevalences
on galaxy mass, led \citet{bes05b} to conclude that low luminosity
radio--loud AGN activity\footnote{The volume probed by the SDSS data is
insufficient to investigate the rare, powerful, classical double
(\citet{fan74} class 2, or FR\,II) radio sources, which do have associated
optical AGN activity, but instead only investigate the distinct population
of lower luminosity (mostly FR\,I) radio sources.} and emission line AGN
activity are distinct physical phenomenon. They argued that emission--line
AGN activity tracks the accretion of cold material onto black holes
through standard accretion disks, and thus reflects the major mode of
growth of the black holes \citep[see discussion in][]{hec04}, whilst low
luminosity radio sources represent re-triggering of pre--formed massive
black holes in old elliptical galaxies. The accretion rates in these radio
sources are comparably low, and the activity is reflected in the
production of low power radio jets but relatively little emission at
optical, ultraviolet and infrared wavelengths. \citet{bes05b} argued that
the accreting material in the radio mode was hot gas, cooling out of the
envelopes of elliptical galaxies. The results for the brightest cluster
galaxies lend further credence to this scenario: since BCGs are located at
the bottom of the cluster potential well, they are also the repository for
cooling intracluster gas. This increased supply of hot gas naturally leads
to the observed increased radio--loud AGN fraction.

\citet{bes06a} suggested that one possibility for the accretion mechanism
is Bondi--Hoyle accretion from a strong cooling flow \citep[see also the
recent work of][]{all06}; the accretion rate for this is given by
\citet{cro06} as $\dot{m}_{\rm Bondi} \approx G \mu m_p k T M_{\rm BH} /
\Lambda$, where $\mu m_p$ is the mean mass of particles in the gas, and
$\Lambda$ is the cooling function. For isothermal elliptical galaxy halos,
$T \propto \sigma^2$; the black hole mass versus velocity dispersion
relation \citep[e.g.][]{tre02} then implies $T \propto M_{\rm
BH}^{0.5}$. $\Lambda$ is relatively independent of temperature (and hence
black hole mass) at the temperature of elliptical galaxy haloes, and
therefore the Bondi accretion rate scales roughly as $M_{\rm BH}^{1.5}$,
which is approximately the same exponent as the radio--loud AGN fraction
for all galaxies. Interestingly, for BCGs the approximation $T \propto
M_{\rm BH}^{0.5}$ is not valid, because the temperature of the gas
accreting on to the BCG is not directly related to the properties of the
host galaxy, but rather is controlled by the larger--scale environment. In
a cluster of given temperature, therefore, the Bondi--Hoyle accretion rate
might be expected to scale linearly with the black hole mass (and hence
stellar mass) of the BCG, which is exactly what is observed in
Figure~\ref{bcgradfrac}.

One surprising result is that there is no increase in the radio--loud AGN
fraction in higher velocity dispersion (mass) clusters: these would be
expected to have higher cooling rates, raising the radio--loud fraction.
For the Bondi--Hoyle mechanism, at fixed black hole mass the accretion
rate scales as $T / \Lambda$, and at the temperature of clusters,
$\Lambda$ increases approximately as $T^{0.5}$. The accretion rate
therefore scales as $T^{0.5}$, and thus roughly linearly with the cluster
velocity dispersion, $\sigma_v$. This would predict nearly a factor of two
difference between the accretion rates, and hence the radio-loud AGN
fractions, of the clusters in the low and high velocity dispersion bins,
which ought to be (just) detectable from the current data.  The failure to
observe this effect may be because the relevant temperature is that of the
gas in the central regions of the cluster, and most massive clusters have
cooler cores associated with the cooling flow; this reduces the expected
difference in accretion rate between high and low mass clusters, possibly
to within the errors of the current measurements.  Investigating this with
a much larger sample of clusters, or using clusters for which the core
temperatures of the X--ray gas have been measured, would provide a
critical test of this model.

The increased likelihood of radio--loud AGN activity for non--BCG galaxies
within 0.2$r_{200}$, particularly the more massive ones, suggests that the
accretion of gas onto these galaxies is also controlled to some extent by
the intracluster medium. It is intriguing that the cooling radius of
clusters (that is, the radius within which the cooling time is less than
the Hubble time) is typically about 10\% of $r_{200}$: thus, the
intracluster medium affects the radio--loud fraction of only those
galaxies within, or close to, the cooling radius. 

The emission--line properties of the BCGs provide further support for the
interpretation that different accreting material is responsible for
low--luminosity radio--loud AGN activity than that for emission line AGN
activity. In contrast to the radio--loud AGN activity, the emission--line
properties of BCGs are in general suppressed relative to galaxies outside
clusters, and are comparable to those of other galaxies of the same mass
in the inner regions of the clusters. This implies that there is a lack of
cold gas to fuel the black holes in these galaxies.  This result is
undoubtedly related to the decreased levels of star formation activity in
cluster environments \citep[e.g.][and references
therein]{lew02,gom03,mat04}.

\begin{figure}
\centerline{
\psfig{file=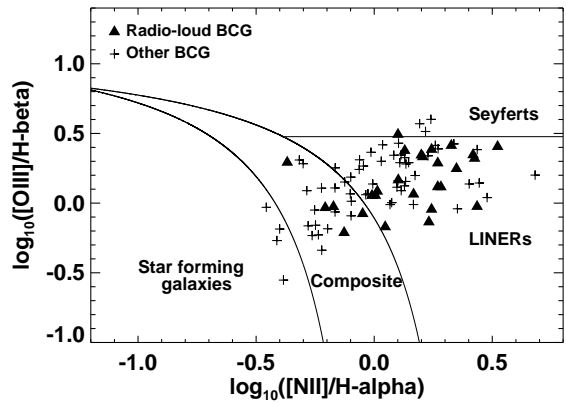,angle=0,width=8.3cm,clip=}
}
\caption{\label{bptplot} An emission line diagnostic plot for the
brightest cluster galaxies with emission lines, divided into different
classifications for the source of ionisation of the emission lines
following \citet{kau03c}. The majority of these BCGs, especially the
radio-loud ones, have LINER-like emission line spectra.}
\end{figure}

Interestingly, those BCGs which do show emission--line activity also show
an enhanced level of radio--loud AGN activity, suggesting that in BCGs (in
contrast to other galaxies) there may be some connection between the
processes which drive the radio and emission--line activity. This is
consistent with the discussion in Section~\ref{optagn} that the BCGs of
cooling flow clusters frequently have associated extended emission--line
nebulae.  The properties of these are similar to those of low--ionisation
nuclear emission--line regions \citep[LINERs; e.g.][]{voi97}, and so may
be mistaken for AGN activity. Thus, the emission lines detected in some
BCGs may not actually arise from AGN activity, but rather from the
cluster--scale cooling flow which fuels the radio--loud AGN
activity. Figure~\ref{bptplot} shows the location of the SDSS
emission--line BCGs on the traditional `BPT' \citep{bal81} emission line
diagnostic diagram. As a guide, this diagram has been divided into regions
\citep[as defined by][]{kau03c} indicating the expected locations of star
forming galaxies, Seyferts, LINERs, and galaxies whose spectrum is a star
formation -- AGN composite. The majority of the emission--line BCGs,
especially those which are radio--loud AGN, are found in the LINER region
of the diagram, and so some of this line emission may indeed be directly
associated with the cooling flows.

\section{Radio--loud AGN heating of cooling flows}
\label{cooling}

\subsection{The mass--dependent heating rate of BCGs}
\label{heatrate}

Radio--loud AGN have relatively short lifetimes ($10^7 - 10^8$ years), and
so the very high fraction of galaxies which host radio--loud AGN (over
30\% at the highest masses) implies that this activity must be constantly
re--triggered. The probability of a galaxy of a given mass hosting a radio
source of given radio luminosity (cf.\ Figure~\ref{radfuncmass}) can
therefore be interpreted, probabilistically, as the fraction of its time
that a galaxy of a given mass spends as a radio source of given
luminosity. Monochromatic radio luminosity represents only a tiny fraction
of the energetic output of a radio source, however, with the mechanical
energy of the radio jets being 2--3 orders of magnitude larger. In order
to determine the time--averaged heating output of radio--loud AGN,
therefore, it is necessary to derive a conversion between the radio
luminosity and mechanical luminosity of the radio sources.

Any such conversion is necessarily approximate because the precise ratio
between mechanical and radio luminosities is both unknown, and varies
throughout the lifetime of a radio source. Nevertheless, it is reasonable
to assume that there may be some statistical average relationship
(e.g. averaging over all ages of a radio outburst), such that a conversion
between the radio and mechanical luminosity functions can still be carried
out at a population level. Such a conversion should be largely independent
of the host galaxy mass (indeed, if it were not, then the similarity of
the radio luminosity shapes for galaxies of different masses would be a
remarkable co-incidence). The mean conversion relation can be determined
empirically using data from the observed bubbles and cavities produced by
radio sources in the intracluster medium of groups and clusters of
galaxies. \citet{bir04} compiled data for the energies, ages and radio
luminosities of these cavities, and \citet{bes06a} used these data to
determine that:

\begin{equation}
\label{eq1}
\frac{L_{\rm mech}}{10^{36}{\rm W}} = (3.0 \pm 0.2)
\left(\frac{L_{\rm 1.4GHz}}{10^{25}{\rm W Hz}^{-1}}\right)^{0.40 \pm
  0.13}.
\end{equation}

Given the result of Section~\ref{radbcg} that the {\it shape} of the radio
luminosity function is the same between galaxies of all masses, then
converting the radio luminosity function to a mechanical luminosity
function using Equation~\ref{eq1} means that the shape of the mechanical
luminosity function is also similar for galaxies of different masses (and
for BCGs). Integrating across this, Best \etal\ estimated the
time--averaged AGN heating output of all galaxies, as a function of black
hole mass\footnote{It is worth noting that \citet{nip05} adopt a similar
method but with the opposite approach: beginning from the assumption that
AGN mechanical heating balances cooling, they use the \citet{bir04}
results to derive a radio luminosity function for radio--loud AGN, which
they show to be in good agreement with that observed.}:

\begin{equation}
\label{eqhall}
\bar{H}_{\rm all} = 1.6 \times 10^{34} (M_{\rm BH} / 10^8 M_{\odot})^{1.6}
{\rm W}.
\end{equation}

In this equation, the black hole mass dependence arises from directly from
the scaling of the radio-loud fraction with black hole mass (ie. that more
massive black holes are switched on for a larger fraction of the
time). Errors in the conversion of radio to mechanical luminosity, and
other such uncertainties, only affect the normalisation factor. This can
be accounted for by introducing a factor $f$ into Equation~\ref{eqhall},
giving:

\begin{equation}
\label{eqhall2}
\bar{H}_{\rm all} = 1.6 \times 10^{34} f (M_{\rm BH} / 10^8 M_{\odot})^{1.6}
{\rm W}.
\end{equation}

The factor $f$ accounts for a range of possible uncertainties, and these
are discussed in detail in Section~\ref{ffactor}. It is likely, however,
that it has a value reasonably close to unity. It is noteworthy that
\citet{bes06a} found that a value $f \approx 1$ was appropriate for
heating to balance radiative cooling losses in elliptical galaxies.

The method described above can also be applied to the brightest cluster
galaxies. Taking the result from Section~\ref{radbcg} that the radio
luminosity function of BCGs has the same shape as that of `all
galaxies'\footnote{Note that if the radio luminosity function for BCGs
shows a weaker break (which would be consistent with the observations for
clusters with $\sigma_{\rm v} > 500$km\,s$^{-1}$) then the derived heating
rate would be higher --- but even if there was no break at all out to an
extreme luminosity of $10^{30}$W\,Hz$^{-1}$ the increase would only be a
factor of 1.7.} and that only the normalisation differs, then integrating
the distributions in Figure~\ref{radfuncmass}, and incorporating the same
factor $f$ gives:

\begin{equation}
\label{eqhbcg1}
\bar{H}_{\rm BCG} = 2.3 \times 10^{35} f (M_* / 10^{11} M_{\odot}) {\rm W}.
\end{equation}

As shown in Section~\ref{radclus}, the time averaged heating rate of the
non--BCG cluster galaxies outwith 0.2$r_{200}$ is the same as that for
`all galaxies', given in Equation~\ref{eqhall2}. For the cluster galaxies
within 0.2$r_{200}$, the boosted radio--loud AGN fraction will produce a
heating rate between that of the BCGs and that of all galaxies, but the
sample is too small to accurately determine this. For the analysis in the
following sections, the limiting cases of $\bar{H}_{\rm all}$ and
$\bar{H}_{\rm BCG}$ were both considered for these galaxies, and it was
found that there was negligible difference between the results produced.

\subsection{The uncertainty factor $f$}
\label{ffactor}

The analysis method adopted here assumes that a good mean conversion can
be made between radio and mechanical luminosity. There are, however,
examples of cluster radio sources \citep[e.g., MS0735.6+7421, Abell
1835;][]{mcn05,mcn06} where the mechanical power estimate is 2--3 orders
of magnitude higher than that estimated by Equation~\ref{eqhbcg1}.
\citet{bir04} and \citet{raf06} showed that the mechanical luminosity also
relates strongly to the X-ray luminosity of the system, suggesting that a
full understanding of these systems will require consideration of more
parameters than simply the radio luminosity. Nevertheless, the uncertainty
factor $f$ in Equations~\ref{eqhall2} and~\ref{eqhbcg1} can be used to
broadly encompass these additional effects. It is therefore worthwhile
considering in detail what uncertainties are contained within the factor
$f$, as this has important consequences for the interpretation of the
results.  The uncertainty can be categorised in three parts.

\begin{enumerate}
\item {\it Uncertainty in the estimates of cavity mechanical
luminosities.} The calculations of B{\^ i}rzan \etal\ have significant
uncertainties in the estimates of both the energies and the ages of the
radio source cavities. In addition, B{\^ i}rzan \etal\ calculate the pV
energy of the cavity, but the enthalpy of the cavity is given by
$\frac{\gamma}{\gamma-1}$pV, which is 4pV for the relativistic plasma of
the radio lobes. There may also be additional heating directly from the
radio jets: \citet{nus06} argue that the mechanical energy may exceed pV
by a factor of 10. 

\item {\it Uncertainty in the radio to mechanical luminosity conversion.}
As discussed above, the varying radio luminosity during a source lifetime
means that this conversion is only accurate in an average sense over the
whole radio source population; however, to first order it ought to be both
reasonable, and largely independent of host galaxy mass. There have been
arguments that this conversion might be dependent upon environment,
however: for fixed jet kinetic power, radio luminosities could be higher
in clusters, due to the confining effect of the dense intracluster medium
reducing adiabatic expansion losses in the radio lobes and therefore
increasing the radio synchrotron emission \citep[e.g.][and references
therein]{bar96a}. If correct, this would lead to slightly lower values of
$f$ in denser systems. Alternatively, as suggested by systems such as
MS0735.6+7421 and Abell 1835 discussed above, individual outbursts with
extreme mechanical to radio luminosity ratios may become more important in
clusters, effectively raising the value of $f$.

\item {\it Uncertainty in how much of the cavity mechanical energy gets
converted to heat within the cooling radius}. The heating rate
calculations above assume that all of the mechanical energy of the
cavities is used to heat the intracluster medium, and that such heating
occurs within the cooling radius of the cluster. If either the radio
source energy is not efficiently converted to heat (e.g.  because of
inefficient production of sound/shock waves), or much of that heating
occurs beyond the cooling radius of the cluster, then this could lead to
values of $f \ll 1$.
\end{enumerate}

\subsection{The importance of the BCG in radio--mode heating}
\label{bcgimport}

\begin{figure}
\centerline{
\psfig{file=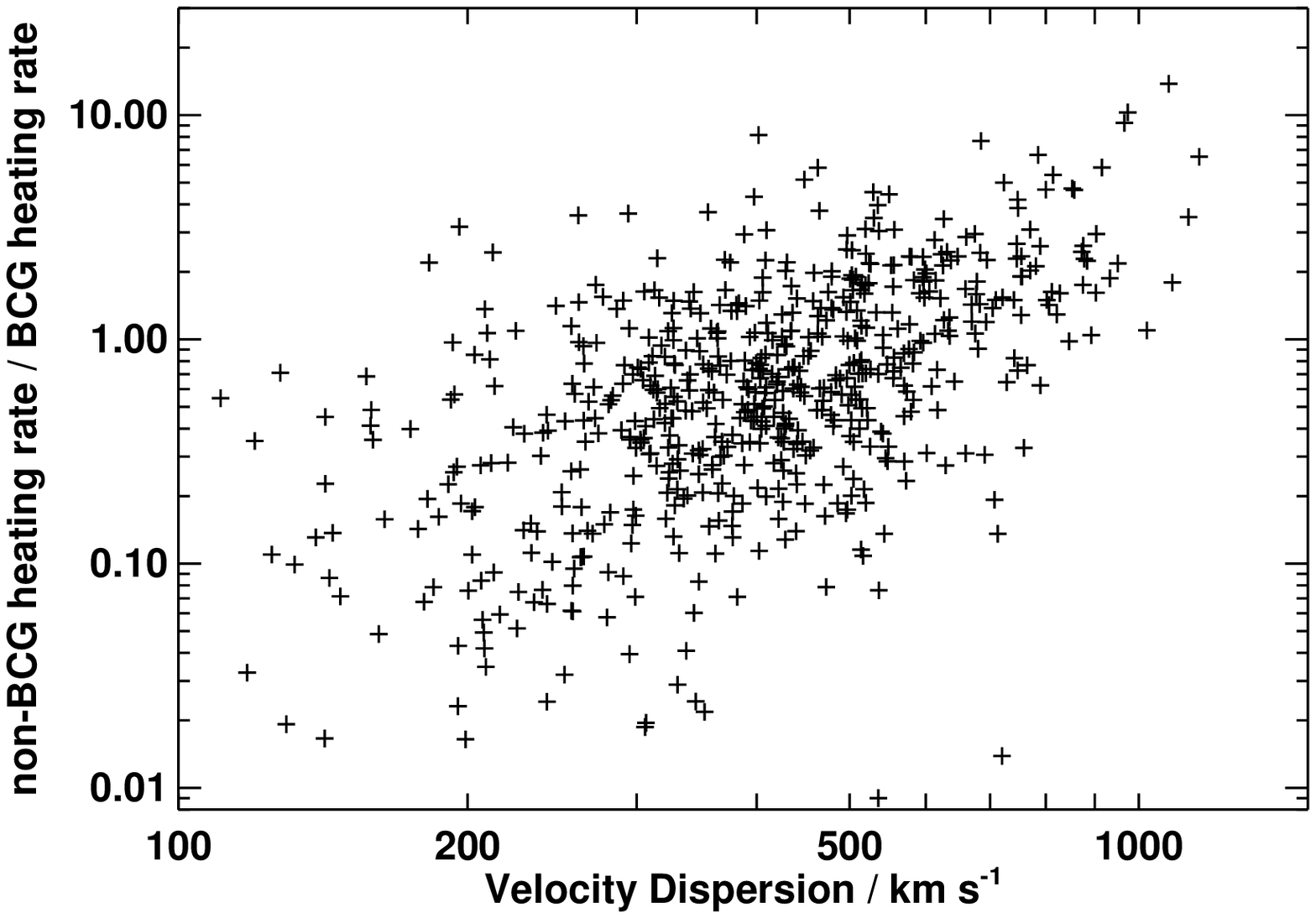,angle=0,width=8.6cm,clip=}
}
\centerline{
\psfig{file=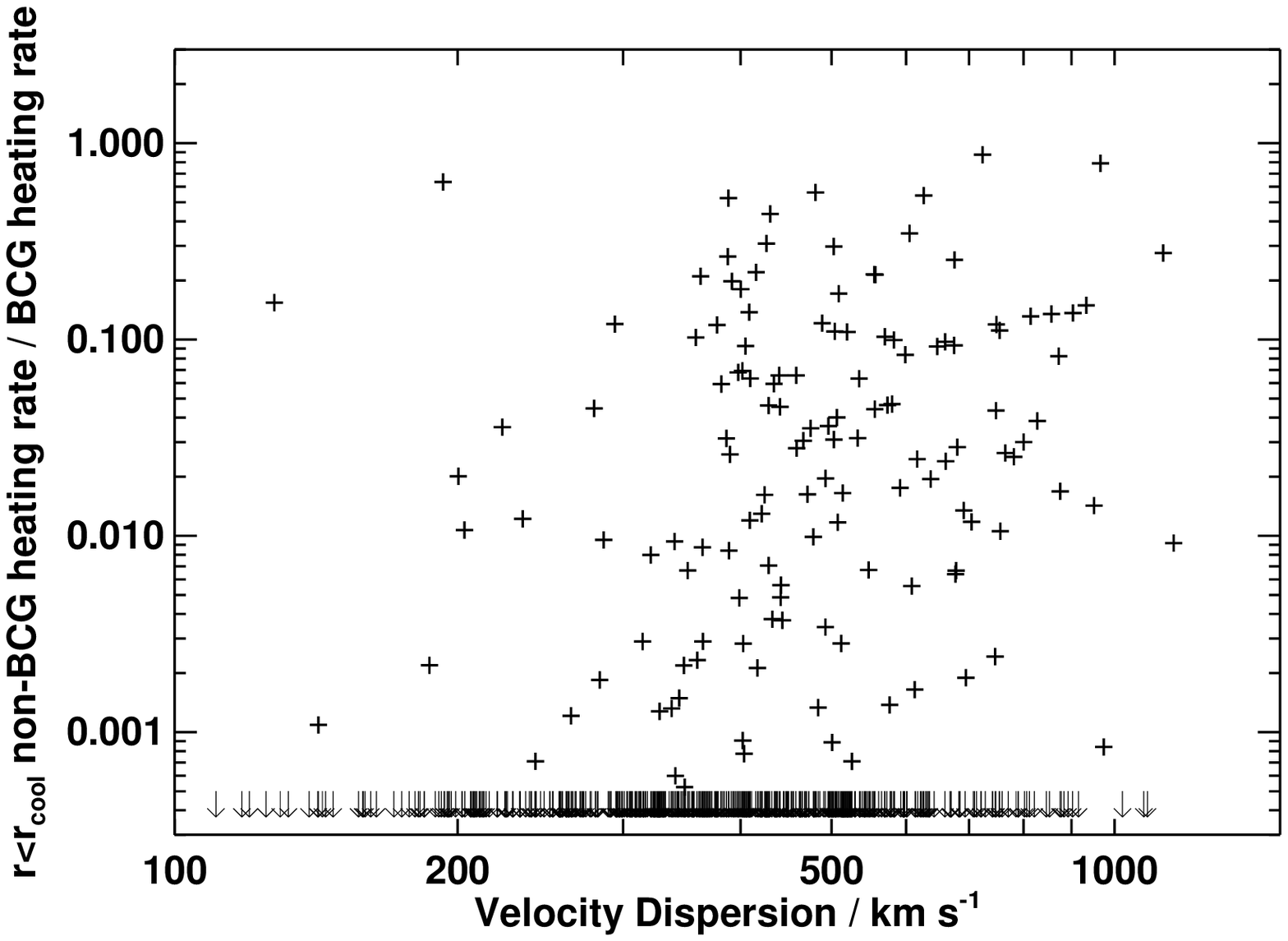,angle=0,width=8.6cm,clip=}
}
\caption{\label{heatcoolbcg} {\it Top:} the ratio of the time--averaged
radio--mode heating rate of all non--BCG cluster galaxies, to that of the
BCG, as a function of velocity dispersion, for the 625 SDSS clusters. {\it
Bottom:} the same plot, but restricted to only those non--BCG cluster
galaxies which are projected to lie within the cooling radius of the
cluster. }
\end{figure}

The top panel of Figure~\ref{heatcoolbcg} shows the ratio of the
time-averaged radio--mode heating rate, summed across all non-BCG cluster
galaxies\footnote{Although the SDSS data only include the more luminous
cluster galaxies, the very strong mass dependence of the radio--loud
fraction means that the missing fainter galaxies should have a negligible
contribution to the radio AGN heating rate. The SDSS data will also miss
some of the brighter non--BCG cluster galaxies, for example due to fibre
collisions, but the number of these will be small.}, to that of the BCG,
for all of the 625 SDSS clusters. These are calculated using
Equations~\ref{eqhall2} and~\ref{eqhbcg1}, assuming that the same value of
$f$ is appropriate for both. In most groups and low mass clusters the
heating is dominated by the BCG alone, but for the richer clusters, the
sum total of other galaxies can provide significantly more heating than
the BCG alone. This heating from the non--BCG galaxies is, however, spread
over a volume very much larger than the cooling radius of the cluster.

\begin{figure}
\centerline{
\psfig{file=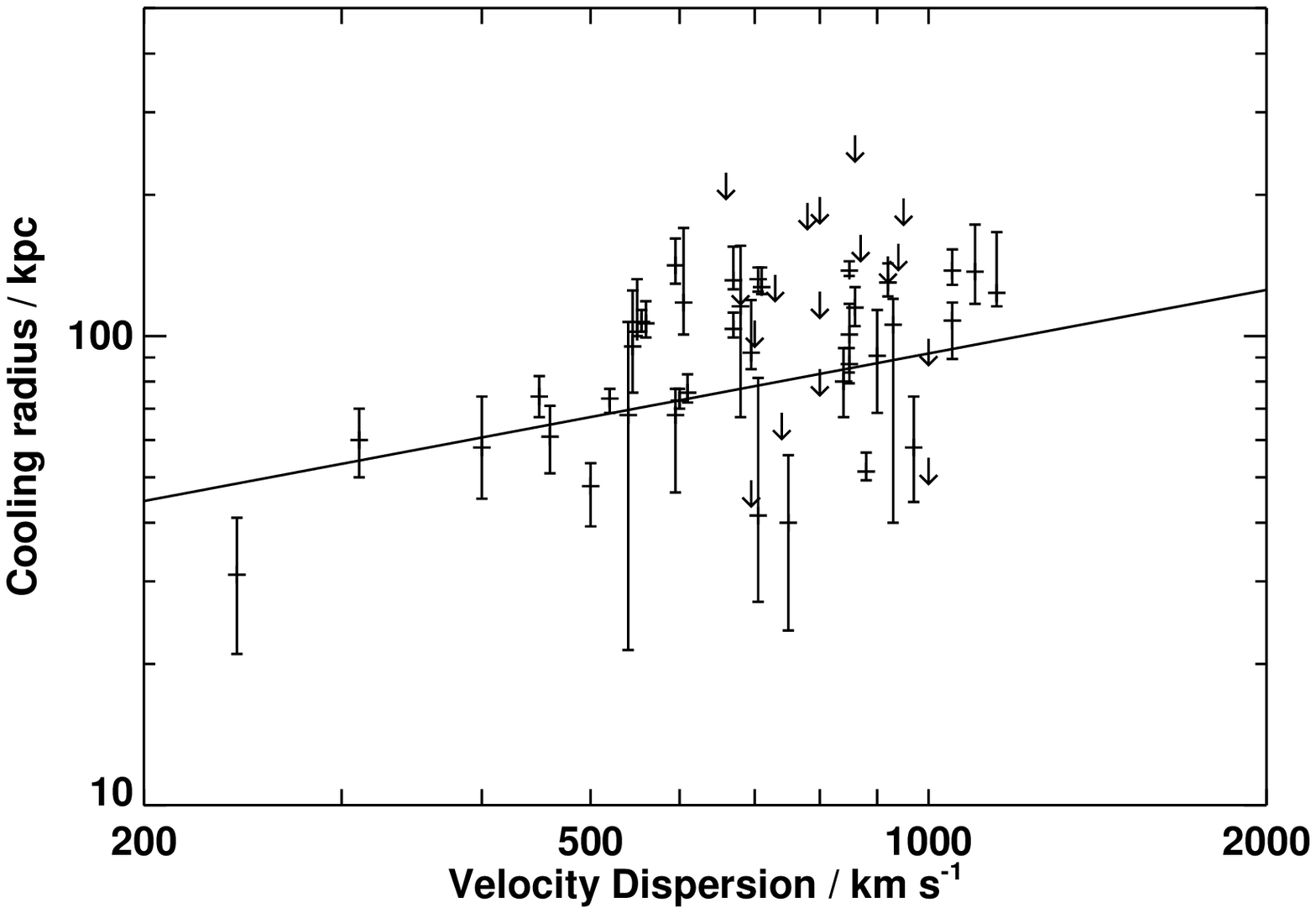,angle=0,width=8.6cm,clip=}
} 
\centerline{
\psfig{file=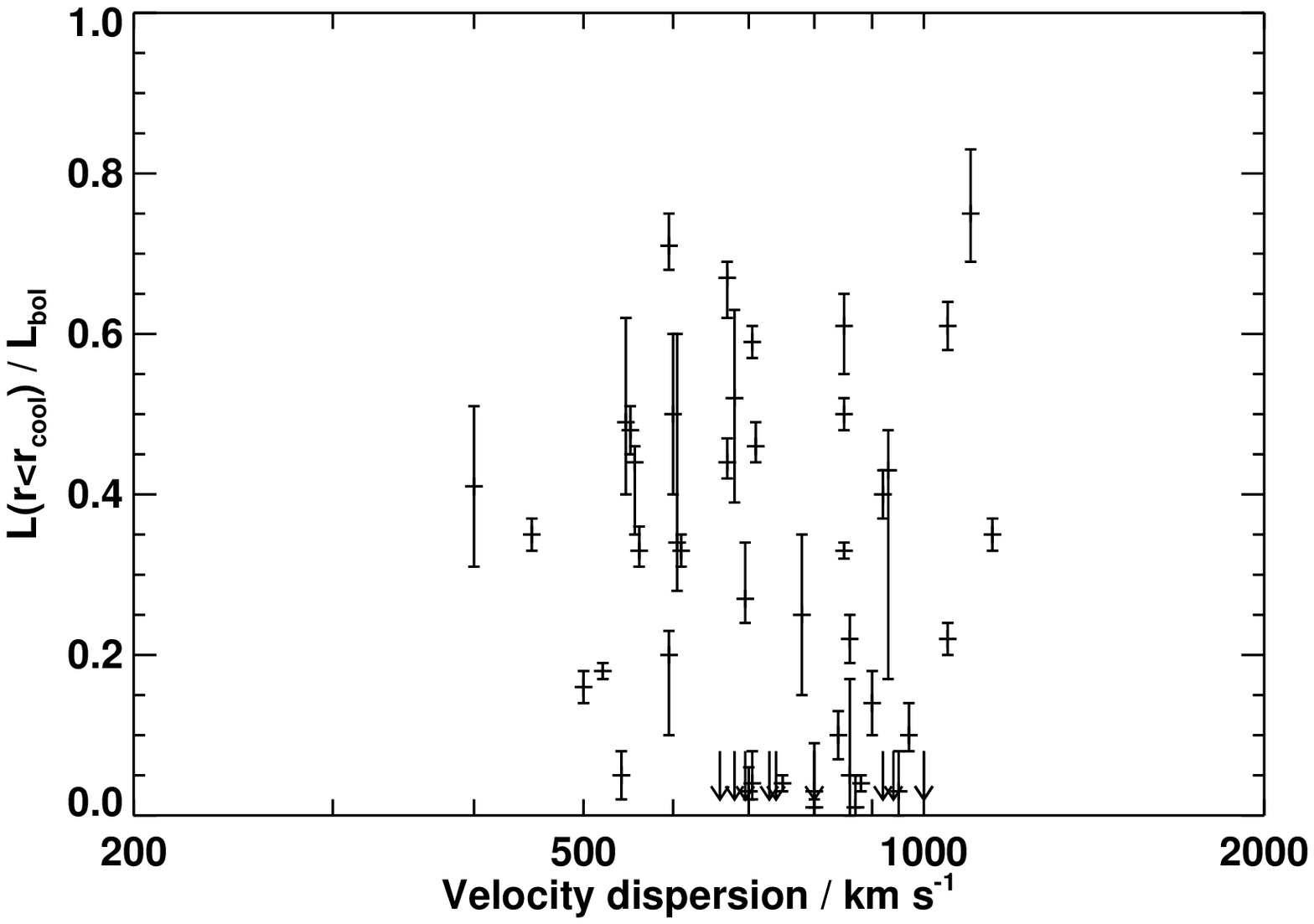,angle=0,width=8.6cm,clip=}}
\caption{\label{rcoolplot} {\it Top:} the observed cooling radii of
clusters and groups of galaxies, taken from \citet{per98}, \citet{sun03},
\citet{xue04} and \citet{kho04}, as a function of cluster velocity
dispersion. The solid line represents a fit to the data using survival
analysis techniques to properly account for the censored data. {\it
Bottom:} the fraction of the X--ray luminosity of those clusters that
arises from within the cooling radius, as calculated by \citet{per98}.}
\end{figure}

\citet{per98} have determined the cooling radii of an X--ray flux limited
sample of 55 nearby clusters, and these cooling radii (converted to the
cosmology adopted here) are plotted as a function of cluster velocity
dispersion in the upper panel of Figure~\ref{rcoolplot}. As most of these
clusters are of high velocity dispersion, these data have been
supplemented by three lower velocity dispersion groups for which cooling
radii have also been measured: NGC1550 \citep{sun03}, RGH80 \citep{xue04}
and NGC6482 \citep{kho04}. The solid line shows a linear fit (in log
space) to the data, calculated using the EM and Buckley--James linear
regression techniques for censored data, within the {\sc ASURV} survival
analysis package \citep{lav92}: $r_{\rm cool} = 67 (\sigma_v /
500$km\,s$^{-1})^{0.45}$kpc. Using this fit to estimate the cooling radius
of each SDSS cluster, the lower panel of Figure~\ref{heatcoolbcg} shows
the ratio of radio--mode heating from non--BCG cluster galaxies projected
within the cooling radius to that of the BCG. This analysis demonstrates
that only the BCG is really relevant when calculating radio--mode heating
within the cooling radius: unless there are other effects not considered
here (e.g. the efficiency factor $f$ varies between BCGs and non-BCGs for
some unknown reason) then the model of \citet{nus06} whereby cooling flows
are suppressed by the combined activity of all cluster galaxies does not
seem to work. This is a result of the very strong mass dependence of the
radio--loud AGN fraction (with BCGs being the highest mass galaxies)
coupled with the additional boosting of the radio--loud fraction amongst
the BCG population.

\subsection{AGN heating versus cooling in clusters}

\begin{figure}
\centerline{
\psfig{file=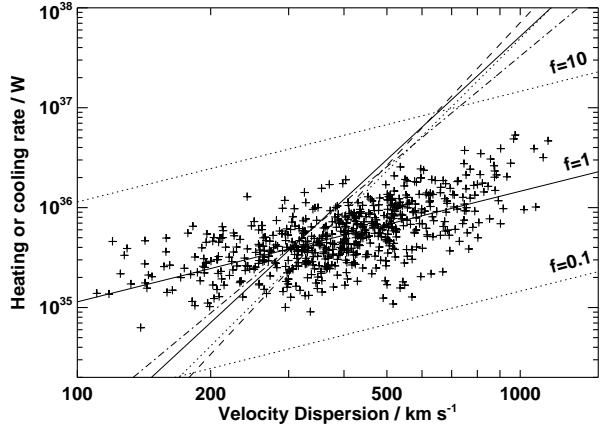,angle=0,width=8.6cm,clip=}
}
\caption{\label{heatcoolsigma} The data points show the time--averaged
heating rate associated with radio--loud AGN activity, assuming $f=1$ in
Equations~\ref{eqhall2} and~\ref{eqhbcg1}, summed across all galaxies in
the clusters, as a function of cluster velocity dispersion. The solid line
through these points, labelled `$f=1$', is a straight-line fit to these
points. The two dotted lines parallel to this represent scaled versions of
this fit, showing where the data points would be located if $f=0.1$ or
$f=10$.  The remaining 4 lines indicate observed relations between
bolometric X--ray luminosity and velocity dispersion for cluster and
groups: solid line --- \citet{ort04}; dotted line --- \citet{mah01};
dashed line --- \citet{xue00}; dash--dot line --- \citet{pop05}.}
\end{figure}

Figure~\ref{heatcoolsigma} shows the time--averaged mechanical luminosity
associated with radio--loud AGN, assuming $f=1$, summed across all cluster
galaxies, as a function of cluster velocity dispersion. For comparison,
four recently derived relations between bolometric X--ray luminosity and
velocity dispersion are also shown \citep{xue00,mah01,ort04,pop05}. The
combined heating rate of all radio--loud AGN increases much more slowly
with cluster velocity dispersion than the $\sigma^4$ dependence of the
radiative cooling rate. It is clear that (for $f=1$) radio--loud AGN
heating falls short of balancing the radiated X--ray luminosity for
essentially all clusters and groups with $\sigma_{\rm v} \gta
300$km\,s$^{-1}$, and by an order of magnitude at the high mass
end. However, much of the X--ray luminosity of the clusters arises from
radii larger than the cooling radius, and does not need to be balanced by
heating in order to avoid catastrophic cooling collapse. It is
instructive, therefore, to calculate what fraction of the X-ray luminosity
within the cooling radius the BCG radio--mode heating is able to balance.

\citet{per98} determined the fraction of the total X--ray luminosity that
is emitted from within the cooling radius for their sample of 55
clusters. These fractions are plotted as a function of velocity dispersion
in the lower panel of Figure~\ref{rcoolplot}. The mean fraction of X--ray
luminosity that arises from within the cooling radius is $\approx$\,25\%,
and there is no significant trend with velocity dispersion. There is
clearly considerable scatter around this mean value, but this is
comparable to the scatter in the calculated AGN heating rates shown in
Figure~\ref{heatcoolsigma}, and so adoption of the mean value is
reasonable. Combining this 25\% fraction with the X--ray luminosity to
velocity dispersion relation from \citet{ort04}\footnote{This is the
middle-most of the four X--ray luminosity to velocity dispersion relations
shown on Figure~\ref{heatcoolsigma}; these relations are fitted to a
combination of (mostly) clusters and (a few) groups. There is no consensus
in the literature as to how the slope of this relation changes for groups,
and so here this mean relation is simply extrapolated down to the group
regime; the conclusions of the paper would be unaffected even if the slope
changed significantly below 300km\,s$^{-1}$ velocity dispersion.} gives
$L_{\rm X}(r<r_{\rm cool}) \approx 1.3 \times 10^{37} (\sigma_{\rm v} /
1000$km\,s$^{-1})^{4.1}$W.

\begin{figure}
\centerline{
\psfig{file=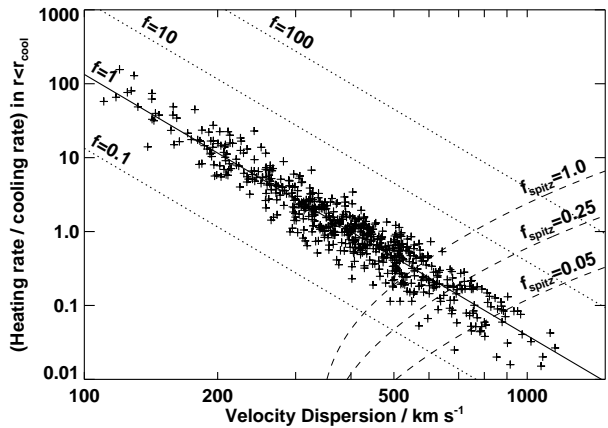,angle=0,width=8.6cm,clip=}
}
\caption{\label{coolfrac} The data points show the ratio of the
time--averaged heating rate associated with radio--loud AGN activity of
all galaxies within the cooling radius, assuming $f=1$ in
Equations~\ref{eqhall2} and~\ref{eqhbcg1}, to the radiative cooling losses
within the cooling radius, as a function of cluster velocity dispersion.
The solid line through these points, labelled `$f=1$', represents the fit
to these points. The dotted lines are scaled versions of this fit, showing
where the data points would be located if $f=0.1$, 10 and 100. Also shown
(dashed lines) are estimates of the amount of heating that can be provided
within the cooling radius by thermal conduction, assuming the radial
temperature distribution of \citet{all01}, and different values of the
suppression factor relative to the maximal Spitzer rate (see
Section~\ref{impli2} for details).}
\end{figure}

Figure~\ref{coolfrac} shows the fraction of the radiative cooling losses
within the cooling radius that can be balanced by AGN heating from
galaxies within that volume, assuming $f=1$. Also shown are the location
that the data points would occupy for alternative values of $f$. One
conclusion is immediately apparent from these results: for a single value
of $f$, radio-mode heating cannot balance radiative cooling losses for all
systems from groups to clusters.

\section{Implications for cooling flows}
\label{impli2}

If radio--loud AGN heating is to exactly balance cooling in all systems,
then the efficiency factor $f$ must increase by a factor of 100--1000
between poor groups and rich clusters. As discussed in
Section~\ref{ffactor}, the factor $f$ includes any uncertainty or
variation in the conversion from radio luminosity to jet kinetic energy,
and also in the efficiency with which the jet kinetic energy is used to
heat the cluster within the cooling radius; either of these factors may be
dependent upon environment. In the former case, any boosting of the radio
luminosities of sources in richer environments due to confinement of the
radio lobes would work in the opposite sense to that required to explain
the trend in Figure~\ref{coolfrac}. The efficiency of converting the radio
source energy into heat, however, offers a more feasible environmental
dependence of $f$. In the most massive clusters it seems quite reasonable
that essentially all of the jet kinetic energy is used to heat the
intracluster medium within the cooling radius of the cluster, since the
cooling radius is large and observations indicate that shock/sound waves
are efficiently produced. In smaller systems, however, the radio source
heating may be more inefficient, either because of inefficient production
of sound/shock waves or because the inflated bubbles rise beyond the
cooling radius before much of their energy is transferred.

A lower efficiency of converting AGN energy into heat in smaller systems
would help towards explaining the trend in Figure~\ref{coolfrac}. However,
given that radio--mode heating is dominated by low luminosity sources
\citep{bes06a}, and such low luminosity sources tend to have small
physical sizes, it seems unlikely that it can account for the required
factor of 100--1000. If it does not, the conclusion has to be that it is
the balance between AGN heating and radiative cooling which changes. In
that case, either radio--loud AGN heating exceeds that required to balance
cooling in groups (but matches in rich clusters; $f \sim 10$), or it falls
short of that required in rich clusters (but matches that in groups; $f
\sim 0.1$), or both ($f \sim 1$).

The possibility that the radio--mode heating in groups exceeds that
required to balance cooling out to $r_{\rm cool}$ is interesting. The
luminosity--temperature relation of groups is observed to have a steep
slope \citep[$L \propto T^5$; e.g.][]{hel00}, much steeper than the slope
of 2 that would be expected from simple scaling laws for virialised gas,
suggesting that the intra-group medium has also been heated by a
non--gravitational energy source. The excess heating required corresponds
to about 1keV per particle, and has been referred to as the `entropy
floor' \citep[e.g.\ ][]{pon99}. One possibility is that the intra-group
gas was pre--heated by supernovae from the early episodes of star
formation in the group galaxies \citep[e.g.][]{kai91,bal99,kay03}.
However, to obtain sufficient energy from supernovae requires an extremely
high efficiency of supernova feedback, and it is also difficult for such
models to maintain consistency with the observed metallicity of that gas
\citep[e.g.][]{wu00,kra00}. An alternative heating source, such as AGN, is
therefore favoured \citep[see the review by][for a full
discussion]{mat03b}, and the overheating of the intragroup medium
suggested here (for $f \gta 0.1$) may provide this. \citet{cro05} found
that the gas in galaxy groups containing radio--loud AGN is typically
hotter than that in radio--quiet groups, providing a direct indication
that radio--source heating can be very significant; more recently,
however, \citet{jet07} found no temperature difference in the very central
regions of groups with and without radio sources, indicating that any
radio source heating occurs preferentially at larger radii.

In order for radio--heating to balance cooling in the richest clusters, an
efficiency factor $f \gta 10$ is required. If all of the cavity enthalpy
is used to heat the intracluster medium, and all of that heat is
transferred within the cooling radius, then a value $f \approx 4$ is
expected. If there is further heating from weak shocks, or errors in the
mechanical luminosity determinations (e.g., significant heating from
sources with extreme mechanical-to-radio luminosity ratios), then a still
higher value could be obtained, and it may be possible to reach the
efficiency required.  Alternatively, an additional source of heating may
be present in the most massive systems. The most likely process to provide
this is thermal conduction, which will work to transport energy from the
reservoir of hot gas outside the cooling radius down into the cooler
cluster centre \citep[e.g.][and references
therein]{nar01,voi02,zak03,voi04}.  The heating rate due to thermal
conduction was originally calculated for a pure hydrogen gas by
\citet{spi62}, and on energy grounds alone thermal conduction at the
maximal Spitzer rate can be sufficient to provide all of the heating
necessary to balance cooling in some clusters \citep{voi04,pop06a}.
However, in the presence of magnetic fields, thermal conduction will be
suppressed; the amount of this suppression remains an open question, but
\citet{nar01} argue that in the presence of turbulent magnetic fields it
may be only a factor of a few. \citet{fab06} support the argument that
thermal conduction is relatively efficient in the inner regions of
clusters, as this is necessary to explain why the shocks seen in the
Perseus cluster are isothermal. Further, \citet{rey05} argue that
viscosity at about 25\% of the Spitzer value will make the radio bubbles
in the intracluster medium stable against Rayleigh--Taylor and
Kelvin--Helmholtz instabilities \citep[although magnetic fields offer
another possibility for providing this required stability;
][]{dey03,kai05}.

Thermal conduction can therefore play an important role in heating galaxy
clusters. \citet{voi04} and \citet{pop06a} show, however, that thermal
conduction cannot provide sufficient heating to balance cooling in all
clusters, and nor does the radial distribution of the heating match that
required. In particular, the temperature gradients in the inner regions of
the clusters are too shallow to provide sufficient inward heat flux at the
cluster centres \citep[see also the discussion of][for the Perseus
cluster]{fab06}.  A combined model of AGN heating in the inner regions of
the cluster, and heating by thermal conduction closer to the cooling
radius, would solve this problem, however. Such a double--heating model
was first developed by \citet{rus02}, and has been shown to provide good
agreement with observed cluster properties \citep[see
also][]{bru03b,hoe04,roy05,fuj05}. The relative importance of thermal
conduction should also increase in more massive clusters \citep{hoe04}, as
would be required to account for the decreasing importance of AGN heating.

\citet{all01} studied the temperature profiles of 6 relaxed clusters
observed with Chandra, and found that all could be reasonable well--fitted
using a universal temperature profile. Assuming that this profile is
appropriate for all clusters\footnote{The clusters studied by
\citet{all01} had velocity dispersions ranging from 800 to nearly
1500\,km\,s$^{-1}$, and so the validity of these calculations for systems
with velocity dispersions much below 800\,km\,s$^{-1}$ is less certain.}
then the total flow of energy to within the cooling radius can be
approximated, for different assumptions of the Spitzer suppression factor,
$f_{\rm Spitz}$. These calculations are detailed in
Appendix~\ref{apptherm}, and the results are displayed on
Figure~\ref{coolfrac}: these confirm that for values of $f_{\rm Spitz}
\approx 0.25$ thermal conduction could provide a large fraction of the
necessary heating in the most massive clusters, but has little importance
for clusters with velocity dispersions below 600--800\,km\,s$^{-1}$.

Finally, it is intriguing that for the expected values of $f \approx 1-4$,
the combination of thermal conduction (with $f_{\rm Spitz} \approx 0.25$)
plus AGN heating approximately balances cooling within the cooling radius
for all systems larger than about 400\,km\,s$^{-1}$. In this scenario, at
lower velocity dispersions the BCG radio--heating would exceed the
radiative cooling losses within the cooling radius, and for systems with
velocity dispersions below about 300\,km\,s$^{-1}$ it would exceed the
radiative cooling losses of the entire intragroup or intracluster medium
(not just that within the cooling radius). At this point, the radio source
activity would lead to heating of the gas, possibly causing some of it to
become unbound. This change in the gas properties may then account for the
change in the luminosity--temperature relation between intragroup and
intracluster gas that occurs at about 300--400\,km\,s$^{-1}$ velocity
dispersion.

\section{Conclusions}
\label{concs}

The conclusions of this work can be summarised as follows:

\begin{itemize}

\item Brightest group or cluster galaxies of all stellar masses are more
likely to host radio--loud AGN than other galaxies of the same mass. The
probability of a BCG hosting a radio--loud AGN scales roughly linearly
with stellar mass, which is shallower than the relation for all
galaxies. BCGs are an order of magnitude more likely to host radio--loud
AGN than other galaxies at stellar masses below $10^{11} M_{\odot}$, but
less than a factor of two more likely at stellar masses above $5 \times
10^{11} M_{\odot}$.

\item The distribution of radio luminosities of BCGs is independent of the
mass of the BCG, and the same as that determined for other galaxies. Only
the normalisation of the radio luminosity function changes, not its shape.

\item The fraction of BCGs of a given stellar mass which host radio--loud
AGN, and the distribution of their radio luminosities, do not depend
strongly on the velocity dispersion of their surrounding group or cluster.

\item Group and cluster galaxies other than the BCG show the same radio
properties as those of field galaxies, except for those within
0.2$r_{200}$ of the centre of the system, which show a boosted likelihood
of being radio--loud.

\item The fraction of galaxies with emission--line AGN activity is smaller
in all group and cluster galaxies within $r_{200}$ than in field galaxies
of the same mass. The suppression increases with decreasing cluster-centric
radius, reaching a factor of 2--3 in the centre. With the exception of
BCGs at the centre of strong cooling flow clusters, BCGs show the same
emission--line AGN fractions as other galaxies of the same mass near the
centre of the group\,/\,cluster.

\item These results support the argument of \citep{bes05b} that
low--luminosity radio--loud AGN activity and emission--line AGN activity
are independent physical phenomenon. It is argued that the radio--loud
activity is associated with the cooling of gas from the hot envelopes of
elliptical galaxies and, in the case of central cluster galaxies, also
from the intracluster medium. Accretion of hot gas from a strong cooling
flow is able to explain both the boosted likelihood of BCGs hosting
radio--loud AGN, and the different slopes of the mass--dependencies of the
radio--AGN fractions for BCGs and other galaxies.

\item Within the cooling radius or a group\,/\,cluster, the mechanical
heating output associated with BCG radio--AGN activity exceeds that of all
other cluster galaxies combined.

\item Either the mechanical--to--radio luminosity ratio or the efficiency
of converting the mechanical energy of the radio source into heating the
intracluster medium must be a factor 100--1000 higher in rich clusters
than in poor groups in order that radio--AGN heating balances radiative
cooling for systems of all masses. If not, then radio--loud AGN heating
either overcompensates the radiative cooling losses in galaxy groups,
providing an explanation for the entropy floor, and\,/\,or falls short of
providing enough heat to counterbalance cooling in the richest
clusters. Thermal conduction could provide the extra energy required in
the richest clusters.
\end{itemize}

\section*{Acknowledgements} 

PNB would like to thank the Royal Society for generous financial support
through its University Research Fellowship scheme. The authors thank the
anonymous referee for a prompt and helpful referee's report. The research
makes use of the SDSS Archive, funding for the creation and distribution
of which was provided by the Alfred P. Sloan Foundation, the Participating
Institutions, the National Aeronautics and Space Administration, the
National Science Foundation, the U.S. Department of Energy, the Japanese
Monbukagakusho, and the Max Planck Society.  The research uses the NVSS
and FIRST radio surveys, carried out using the National Radio Astronomy
Observatory Very Large Array: NRAO is operated by Associated Universities
Inc., under co-operative agreement with the National Science Foundation.

\bibliography{pnb} 

\begin{thebibliography}{}

\bibitem[\protect\citeauthoryear{{Adelman-McCarthy et~al.}}{{Adelman-McCarthy
  et~al.}}{2006}]{ade06}
{Adelman-McCarthy J.~K. et~al.} 2006, ApJ Supp., 162, 38

\bibitem[\protect\citeauthoryear{{Allen}, {Dunn}, {Fabian}, {Taylor} \&
  {Reynolds}}{{Allen} et~al.}{2006}]{all06}
{Allen} S.~W.,  {Dunn} R.~J.~H.,  {Fabian} A.~C.,  {Taylor} G.~B.,
  {Reynolds} C.~S.,  2006, MNRAS, 372, 21

\bibitem[\protect\citeauthoryear{{Allen}, {Schmidt} \& {Fabian}}{{Allen}
  et~al.}{2001}]{all01}
{Allen} S.~W.,  {Schmidt} R.~W.,    {Fabian} A.~C.,  2001, MNRAS, 328, L37

\bibitem[\protect\citeauthoryear{{B{\^ i}rzan}, {Rafferty}, {McNamara}, {Wise}
  \& {Nulsen}}{{B{\^ i}rzan} et~al.}{2004}]{bir04}
{B{\^ i}rzan} L.,  {Rafferty} D.~A.,  {McNamara} B.~R.,  {Wise} M.~W.,
  {Nulsen} P.~E.~J.,  2004, ApJ, 607, 800

\bibitem[\protect\citeauthoryear{Baldwin, Phillips \& Terlevich}{Baldwin
  et~al.}{1981}]{bal81}
Baldwin J.~A.,  Phillips M.~M.,    Terlevich R.,  1981, PASP, 93, 5

\bibitem[\protect\citeauthoryear{{Balogh}, {Babul} \& {Patton}}{{Balogh}
  et~al.}{1999}]{bal99}
{Balogh} M.~L.,  {Babul} A.,    {Patton} D.~R.,  1999, MNRAS, 307, 463

\bibitem[\protect\citeauthoryear{Barthel \& Arnaud}{Barthel \&
  Arnaud}{1996}]{bar96a}
Barthel P.~D.,  Arnaud K.~A.,  1996, MNRAS, 283, L45

\bibitem[\protect\citeauthoryear{Becker, White \& Helfand}{Becker
  et~al.}{1995}]{bec95}
Becker R.~H.,  White R.~L.,    Helfand D.~J.,  1995, ApJ, 450, 559

\bibitem[\protect\citeauthoryear{Benson, Bower, Frenk, Lacey, Baugh \&
  Cole}{Benson et~al.}{2003}]{ben03}
Benson A.~J.,  Bower R.~G.,  Frenk C.~S.,  Lacey C.~G.,  Baugh C.~M.,    Cole
  S.,  2003, ApJ, 599, 38

\bibitem[\protect\citeauthoryear{Bernardi, Hyde, Sheth, Miller \&
  Nichol}{Bernardi et~al.}{2006}]{ber06}
Bernardi M.,  Hyde J.~B.,  Sheth R.~K.,  Miller C.~J.,    Nichol R.~C.,  2006,
  AJ submitted; astro-ph/0607117

\bibitem[\protect\citeauthoryear{{Bernstein} \& {Bhavsar}}{{Bernstein} \&
  {Bhavsar}}{2001}]{ber01}
{Bernstein} J.~P.,  {Bhavsar} S.~P.,  2001, MNRAS, 322, 625

\bibitem[\protect\citeauthoryear{Best}{Best}{2004}]{bes04a}
Best P.~N.,  2004, MNRAS, 351, 70

\bibitem[\protect\citeauthoryear{Best, Kaiser, Heckman \& Kauffmann}{Best
  et~al.}{2006}]{bes06a}
Best P.~N.,  Kaiser C.~R.,  Heckman T.~M.,    Kauffmann G.,  2006, MNRAS, 368,
  L67

\bibitem[\protect\citeauthoryear{Best, Kauffmann, Heckman, Brinchmann, Charlot,
  \v{Z}. Ivezi{\'c} \& White}{Best et~al.}{2005}]{bes05b}
Best P.~N.,  Kauffmann G.,  Heckman T.~M.,  Brinchmann J.,  Charlot S.,  \v{Z}.
  Ivezi{\'c}   White S. D.~M.,  2005, MNRAS, 362, 25

\bibitem[\protect\citeauthoryear{Best, Kauffmann, Heckman \& \v{Z}.
  Ivezi{\'c}}{Best et~al.}{2005}]{bes05a}
Best P.~N.,  Kauffmann G.,  Heckman T.~M.,    \v{Z}. Ivezi{\'c} 2005, MNRAS,
  362, 9

\bibitem[\protect\citeauthoryear{{Blanton}, {Sarazin}, {McNamara} \&
  {Wise}}{{Blanton} et~al.}{2001}]{bla01}
{Blanton} E.~L.,  {Sarazin} C.~L.,  {McNamara} B.~R.,    {Wise} M.~W.,  2001,
  ApJ, 558, L15

\bibitem[\protect\citeauthoryear{{Blanton} \& {Roweis}}{{Blanton} \&
  {Roweis}}{2007}]{bla07}
{Blanton} M.~R.,  {Roweis} S.,  2007, AJ, 133, 734

\bibitem[\protect\citeauthoryear{{Blanton et~al.}}{{Blanton
  et~al.}}{2003}]{bla03}
{Blanton M.~R. et~al.} 2003, AJ, 125, 2348

\bibitem[\protect\citeauthoryear{B{\"o}hringer, Voges, Fabian, Edge \&
  Neumann}{B{\"o}hringer et~al.}{1993}]{boh93}
B{\"o}hringer H.,  Voges W.,  Fabian A.~C.,  Edge A.~C.,    Neumann D.~M.,
  1993, MNRAS, 264, L25

\bibitem[\protect\citeauthoryear{Bower, Benson, Malbon, Helly, Frenk, Baugh,
  Cole \& Lacey}{Bower et~al.}{2006}]{bow06}
Bower R.~G.,  Benson A.~J.,  Malbon R.,  Helly J.~C.,  Frenk C.~S.,  Baugh
  C.~M.,  Cole S.,    Lacey C.~G.,  2006, MNRAS, 370, 645

\bibitem[\protect\citeauthoryear{Brighenti \& Mathews}{Brighenti \&
  Mathews}{2006}]{bri06}
Brighenti F.,  Mathews W.~G.,  2006, ApJ, 643, 120

\bibitem[\protect\citeauthoryear{{Brough}, {Collins}, {Burke}, {Lynam} \&
  {Mann}}{{Brough} et~al.}{2005}]{bro05}
{Brough} S.,  {Collins} C.~A.,  {Burke} D.~J.,  {Lynam} P.~D.,    {Mann} R.~G.,
   2005, MNRAS, 364, 1354

\bibitem[\protect\citeauthoryear{{Br{\"u}ggen}}{{Br{\"u}ggen}}{2003}]{bru03b}
{Br{\"u}ggen} M.,  2003, ApJ, 593, 700

\bibitem[\protect\citeauthoryear{{Br{\"u}ggen} \& {Kaiser}}{{Br{\"u}ggen} \&
  {Kaiser}}{2002}]{bru02}
{Br{\"u}ggen} M.,  {Kaiser} C.~R.,  2002, Nat, 418, 301

\bibitem[\protect\citeauthoryear{{Br{\"u}ggen}, {Ruszkowski} \&
  {Hallman}}{{Br{\"u}ggen} et~al.}{2005}]{bru05}
{Br{\"u}ggen} M.,  {Ruszkowski} M.,    {Hallman} E.,  2005, ApJ, 630, 740

\bibitem[\protect\citeauthoryear{{Burns}}{{Burns}}{1990}]{bur90}
{Burns} J.~O.,  1990, AJ, 99, 14

\bibitem[\protect\citeauthoryear{{Burns}, {White} \& {Hough}}{{Burns}
  et~al.}{1981}]{bur81}
{Burns} J.~O.,  {White} R.~A.,    {Hough} D.~H.,  1981, AJ, 86, 1

\bibitem[\protect\citeauthoryear{Carilli, Perley \& Harris}{Carilli
  et~al.}{1994}]{car94b}
Carilli C.~L.,  Perley R.~A.,    Harris D.~E.,  1994, MNRAS, 270, 173

\bibitem[\protect\citeauthoryear{Cattaneo, Dekel, Devriendt, Guiderdoni \&
  Blaizot}{Cattaneo et~al.}{2006}]{cat06}
Cattaneo A.,  Dekel A.,  Devriendt J.,  Guiderdoni B.,    Blaizot J.,  2006,
  MNRAS, 370, 1651

\bibitem[\protect\citeauthoryear{{Churazov}, {Br{\"u}ggen}, {Kaiser},
  {B{\"o}hringer} \& {Forman}}{{Churazov} et~al.}{2001}]{chu01}
{Churazov} E.,  {Br{\"u}ggen} M.,  {Kaiser} C.~R.,  {B{\"o}hringer} H.,
  {Forman} W.,  2001, ApJ, 554, 261

\bibitem[\protect\citeauthoryear{Condon, Cotton, Greisen, Yin, Perley, Taylor
  \& Broderick}{Condon et~al.}{1998}]{con98}
Condon J.~J.,  Cotton W.~D.,  Greisen E.~W.,  Yin Q.~F.,  Perley R.~A.,  Taylor
  G.~B.,    Broderick J.~J.,  1998, AJ, 115, 1693

\bibitem[\protect\citeauthoryear{{Croston}, {Hardcastle} \&
  {Birkinshaw}}{{Croston} et~al.}{2005}]{cro05}
{Croston} J.~H.,  {Hardcastle} M.~J.,    {Birkinshaw} M.,  2005, MNRAS, 357,
  279

\bibitem[\protect\citeauthoryear{{Croton et~al.}}{{Croton
  et~al.}}{2006}]{cro06}
{Croton D. et~al.} 2006, MNRAS, 365, 11

\bibitem[\protect\citeauthoryear{{Dalla Vecchia}, {Bower}, {Theuns}, {Balogh},
  {Mazzotta} \& {Frenk}}{{Dalla Vecchia} et~al.}{2004}]{dal04}
{Dalla Vecchia} C.,  {Bower} R.~G.,  {Theuns} T.,  {Balogh} M.~L.,  {Mazzotta}
  P.,    {Frenk} C.~S.,  2004, MNRAS, 355, 995

\bibitem[\protect\citeauthoryear{{David}, {Nulsen}, {McNamara}, {Forman},
  {Jones}, {Ponman}, {Robertson} \& {Wise}}{{David} et~al.}{2001}]{dav01}
{David} L.~P.,  {Nulsen} P.~E.~J.,  {McNamara} B.~R.,  {Forman} W.,  {Jones}
  C.,  {Ponman} T.,  {Robertson} B.,    {Wise} M.,  2001, ApJ, 557, 546

\bibitem[\protect\citeauthoryear{{De Lucia} \& Blaizot}{{De Lucia} \&
  Blaizot}{2006}]{del06}
{De Lucia} G.,  Blaizot J.,  2006, MNRAS, submitted; astro-ph/0606519

\bibitem[\protect\citeauthoryear{{De Young}}{{De Young}}{2003}]{dey03}
{De Young} D.~S.,  2003, MNRAS, 343, 719

\bibitem[\protect\citeauthoryear{{Dunn}, {Fabian} \& {Taylor}}{{Dunn}
  et~al.}{2005}]{dun05}
{Dunn} R.~J.~H.,  {Fabian} A.~C.,    {Taylor} G.~B.,  2005, MNRAS, 364, 1343

\bibitem[\protect\citeauthoryear{{Edge}}{{Edge}}{1991}]{edg91}
{Edge} A.~C.,  1991, MNRAS, 250, 103

\bibitem[\protect\citeauthoryear{Edwards, Hudson, Balogh \& Smith}{Edwards
  et~al.}{2007}]{edw07}
Edwards L. O.~V.,  Hudson M.~J.,  Balogh M.~L.,    Smith R.~J.,  2007, MNRAS,
  submitted

\bibitem[\protect\citeauthoryear{Fabian}{Fabian}{1994}]{fab94}
Fabian A.~C.,  1994, ARA\&A, 32, 277

\bibitem[\protect\citeauthoryear{Fabian, Sanders, Allen, Crawford, Iwasawa,
  Johnstone, Schmidt \& Taylor}{Fabian et~al.}{2003}]{fab03}
Fabian A.~C.,  Sanders J.~S.,  Allen S.~W.,  Crawford C.~S.,  Iwasawa K.,
  Johnstone R.~M.,  Schmidt R.~W.,    Taylor G.~B.,  2003, MNRAS, 344, L43

\bibitem[\protect\citeauthoryear{{Fabian}, {Sanders}, {Taylor}, {Allen},
  {Crawford}, {Johnstone} \& {Iwasawa}}{{Fabian} et~al.}{2006}]{fab06}
{Fabian} A.~C.,  {Sanders} J.~S.,  {Taylor} G.~B.,  {Allen} S.~W.,  {Crawford}
  C.~S.,  {Johnstone} R.~M.,    {Iwasawa} K.,  2006, MNRAS, 366, 417

\bibitem[\protect\citeauthoryear{{Fabian et~al.}}{{Fabian
  et~al.}}{2000}]{fab00b}
{Fabian A.~C. et~al.} 2000, MNRAS, 318, L65

\bibitem[\protect\citeauthoryear{Fanaroff \& Riley}{Fanaroff \&
  Riley}{1974}]{fan74}
Fanaroff B.~L.,  Riley J.~M.,  1974, MNRAS, 167, 31P

\bibitem[\protect\citeauthoryear{{Finn}, {Zaritsky}, {McCarthy} Jr.,
  {Poggianti}, {Rudnick}, {Halliday}, {Milvang-Jensen}, {Pell{\'o}} \&
  {Simard}}{{Finn} et~al.}{2005}]{fin05}
{Finn} R.~A.,  {Zaritsky} D.,  {McCarthy} Jr. D.~W.,  {Poggianti} B.,
  {Rudnick} G.,  {Halliday} C.,  {Milvang-Jensen} B.,  {Pell{\'o}} R.,
  {Simard} L.,  2005, ApJ, 630, 206

\bibitem[\protect\citeauthoryear{{Forman et~al.}}{{Forman
  et~al.}}{2005}]{for05}
{Forman W. et~al.} 2005, ApJ, 635, 894

\bibitem[\protect\citeauthoryear{{Forman et~al.}}{{Forman
  et~al.}}{2006}]{for06}
{Forman W. et~al.} 2006, ApJ submitted; astro-ph/0604583

\bibitem[\protect\citeauthoryear{{Fujita} \& {Suzuki}}{{Fujita} \&
  {Suzuki}}{2005}]{fuj05}
{Fujita} Y.,  {Suzuki} T.~K.,  2005, ApJ, 630, L1

\bibitem[\protect\citeauthoryear{{G{\' o}mez et~al.}}{{G{\' o}mez
  et~al.}}{2003}]{gom03}
{G{\' o}mez et~al.} 2003, ApJ, 584, 210

\bibitem[\protect\citeauthoryear{{Hatch}, {Crawford}, {Johnstone} \&
  {Fabian}}{{Hatch} et~al.}{2006}]{hat06}
{Hatch} N.~A.,  {Crawford} C.~S.,  {Johnstone} R.~M.,    {Fabian} A.~C.,  2006,
  MNRAS, 367, 433

\bibitem[\protect\citeauthoryear{Heckman, Baum, {van Breugel} \&
  McCarthy}{Heckman et~al.}{1989}]{hec89}
Heckman T.~M.,  Baum S.~A.,  {van Breugel} W. J.~M.,    McCarthy P.~J.,  1989,
  ApJ, 338, 48

\bibitem[\protect\citeauthoryear{Heckman, Kauffmann, Brinchmann, Charlot,
  Tremonti \& White}{Heckman et~al.}{2004}]{hec04}
Heckman T.~M.,  Kauffmann G.,  Brinchmann J.,  Charlot S.,  Tremonti C.,
  White S.~D.,  2004, ApJ, 613, 109

\bibitem[\protect\citeauthoryear{{Helsdon} \& {Ponman}}{{Helsdon} \&
  {Ponman}}{2000}]{hel00}
{Helsdon} S.~F.,  {Ponman} T.~J.,  2000, MNRAS, 315, 356

\bibitem[\protect\citeauthoryear{Hill \& Lilly}{Hill \& Lilly}{1991}]{hil91}
Hill G.~J.,  Lilly S.~J.,  1991, ApJ, 367, 1

\bibitem[\protect\citeauthoryear{{Hoeft} \& {Br{\"u}ggen}}{{Hoeft} \&
  {Br{\"u}ggen}}{2004}]{hoe04}
{Hoeft} M.,  {Br{\"u}ggen} M.,  2004, ApJ, 617, 896

\bibitem[\protect\citeauthoryear{{Jetha}, {Ponman}, {Hardcastle} \&
  {Croston}}{{Jetha} et~al.}{2007}]{jet07}
{Jetha} N.~N.,  {Ponman} T.~J.,  {Hardcastle} M.~J.,    {Croston} J.~H.,  2007,
  MNRAS, 376, 193

\bibitem[\protect\citeauthoryear{{Kaastra}, {Ferrigno}, {Tamura}, {Paerels},
  {Peterson} \& {Mittaz}}{{Kaastra} et~al.}{2001}]{kaa01}
{Kaastra} J.~S.,  {Ferrigno} C.,  {Tamura} T.,  {Paerels} F.~B.~S.,  {Peterson}
  J.~R.,    {Mittaz} J.~P.~D.,  2001, A\&A, 365, L99

\bibitem[\protect\citeauthoryear{{Kaiser} \& {Binney}}{{Kaiser} \&
  {Binney}}{2003}]{kai03}
{Kaiser} C.~R.,  {Binney} J.,  2003, MNRAS, 338, 837

\bibitem[\protect\citeauthoryear{{Kaiser}, {Pavlovski}, {Pope} \&
  {Fangohr}}{{Kaiser} et~al.}{2005}]{kai05}
{Kaiser} C.~R.,  {Pavlovski} G.,  {Pope} E.~C.~D.,    {Fangohr} H.,  2005,
  MNRAS, 359, 493

\bibitem[\protect\citeauthoryear{{Kaiser}}{{Kaiser}}{1991}]{kai91}
{Kaiser} N.,  1991, ApJ, 383, 104

\bibitem[\protect\citeauthoryear{Kauffmann, White, Heckman, M{\'e}nard,
  Brinchmann, Charlot, Tremonti \& Brinkmann}{Kauffmann et~al.}{2004}]{kau04}
Kauffmann G.,  White S. D.~M.,  Heckman T.~M.,  M{\'e}nard B.,  Brinchmann J.,
  Charlot S.,  Tremonti C.,    Brinkmann J.,  2004, MNRAS, 353, 713

\bibitem[\protect\citeauthoryear{{Kauffmann et~al.}}{{Kauffmann
  et~al.}}{2003a}]{kau03a}
{Kauffmann G. et~al.} 2003a, MNRAS, 341, 33

\bibitem[\protect\citeauthoryear{{Kauffmann et~al.}}{{Kauffmann
  et~al.}}{2003b}]{kau03b}
{Kauffmann G. et~al.} 2003b, MNRAS, 341, 54

\bibitem[\protect\citeauthoryear{{Kauffmann et~al.}}{{Kauffmann
  et~al.}}{2003c}]{kau03c}
{Kauffmann G. et~al.} 2003c, MNRAS, 346, 1055

\bibitem[\protect\citeauthoryear{{Kay}, {Thomas} \& {Theuns}}{{Kay}
  et~al.}{2003}]{kay03}
{Kay} S.~T.,  {Thomas} P.~A.,    {Theuns} T.,  2003, MNRAS, 343, 608

\bibitem[\protect\citeauthoryear{{Khosroshahi}, {Jones} \&
  {Ponman}}{{Khosroshahi} et~al.}{2004}]{kho04}
{Khosroshahi} H.~G.,  {Jones} L.~R.,    {Ponman} T.~J.,  2004, MNRAS, 349, 1240

\bibitem[\protect\citeauthoryear{{Kravtsov} \& {Yepes}}{{Kravtsov} \&
  {Yepes}}{2000}]{kra00}
{Kravtsov} A.~V.,  {Yepes} G.,  2000, MNRAS, 318, 227

\bibitem[\protect\citeauthoryear{{Lauer et~al.}}{{Lauer et~al.}}{2006}]{lau06}
{Lauer T.~R. et~al.} 2006, ApJ submitted; astro-ph/0606739

\bibitem[\protect\citeauthoryear{LaValley, Isobe \& Feigelson}{LaValley
  et~al.}{1992}]{lav92}
LaValley M.,  Isobe T.,    Feigelson E.,  1992, BAAS, 24, 839

\bibitem[\protect\citeauthoryear{Ledlow \& Owen}{Ledlow \& Owen}{1996}]{led96}
Ledlow M.~J.,  Owen F.~N.,  1996, AJ, 112, 9

\bibitem[\protect\citeauthoryear{{Lewis et~al.}}{{Lewis et~al.}}{2002}]{lew02}
{Lewis I. et~al.} 2002, MNRAS, 334, 673

\bibitem[\protect\citeauthoryear{{Lin} \& {Mohr}}{{Lin} \&
  {Mohr}}{2004}]{lin04}
{Lin} Y.-T.,  {Mohr} J.~J.,  2004, ApJ, 617, 879

\bibitem[\protect\citeauthoryear{Mahdavi \& Geller}{Mahdavi \&
  Geller}{2001}]{mah01}
Mahdavi A.,  Geller M.~J.,  2001, ApJ, 554, L129

\bibitem[\protect\citeauthoryear{Mateus \& Sodr{\'e}}{Mateus \&
  Sodr{\'e}}{2004}]{mat04}
Mateus A.,  Sodr{\'e} L.,  2004, MNRAS, 349, 1251

\bibitem[\protect\citeauthoryear{Mathews \& Brighenti}{Mathews \&
  Brighenti}{2003}]{mat03b}
Mathews W.~G.,  Brighenti F.,  2003, ARA\&A, 41, 191

\bibitem[\protect\citeauthoryear{{Mathews}, {Faltenbacher} \&
  {Brighenti}}{{Mathews} et~al.}{2006}]{mat06}
{Mathews} W.~G.,  {Faltenbacher} A.,    {Brighenti} F.,  2006, ApJ, 638, 659

\bibitem[\protect\citeauthoryear{{McNamara}, {Nulsen}, {Wise}, {Rafferty},
  {Carilli}, {Sarazin} \& {Blanton}}{{McNamara} et~al.}{2005}]{mcn05}
{McNamara} B.~R.,  {Nulsen} P.~E.~J.,  {Wise} M.~W.,  {Rafferty} D.~A.,
  {Carilli} C.,  {Sarazin} C.~L.,    {Blanton} E.~L.,  2005, Nature, 433, 45

\bibitem[\protect\citeauthoryear{{McNamara et~al.}}{{McNamara
  et~al.}}{2000}]{mcn00}
{McNamara B.~R. et~al.} 2000, ApJ, 534, L135

\bibitem[\protect\citeauthoryear{{McNamara et~al.}}{{McNamara
  et~al.}}{2006}]{mcn06}
{McNamara B.~R. et~al.} 2006, ApJ, 648, 164

\bibitem[\protect\citeauthoryear{Miller, Nichol, Gom{\'e}z, Hopkins \&
  Bernardi}{Miller et~al.}{2003}]{mil03a}
Miller C.~J.,  Nichol R.~C.,  Gom{\'e}z P.~L.,  Hopkins A.~M.,    Bernardi M.,
  2003, ApJ, 597, 142

\bibitem[\protect\citeauthoryear{Miller, Ledlow, Owen \& Hill}{Miller
  et~al.}{2002}]{mil02b}
Miller N.~A.,  Ledlow M.~J.,  Owen F.~N.,    Hill J.~M.,  2002, AJ, 123, 3018

\bibitem[\protect\citeauthoryear{{Miller et~al.}}{{Miller
  et~al.}}{2005}]{mil05}
{Miller C.~J. et~al.} 2005, AJ, 130, 968

\bibitem[\protect\citeauthoryear{{Narayan} \& {Medvedev}}{{Narayan} \&
  {Medvedev}}{2001}]{nar01}
{Narayan} R.,  {Medvedev} M.~V.,  2001, ApJ, 562, L129

\bibitem[\protect\citeauthoryear{{Nipoti} \& {Binney}}{{Nipoti} \&
  {Binney}}{2005}]{nip05}
{Nipoti} C.,  {Binney} J.,  2005, MNRAS, 361, 428

\bibitem[\protect\citeauthoryear{{Nusser}, {Silk} \& {Babul}}{{Nusser}
  et~al.}{2006}]{nus06}
{Nusser} A.,  {Silk} J.,    {Babul} A.,  2006, astro-ph/0602566

\bibitem[\protect\citeauthoryear{Oemler}{Oemler}{1976}]{oem76}
Oemler A.,  1976, ApJ, 209, 693

\bibitem[\protect\citeauthoryear{{Omma}, {Binney}, {Bryan} \& {Slyz}}{{Omma}
  et~al.}{2004}]{omm04}
{Omma} H.,  {Binney} J.,  {Bryan} G.,    {Slyz} A.,  2004, MNRAS, 348, 1105

\bibitem[\protect\citeauthoryear{{Ortiz-Gil}, {Guzzo}, {Schuecker},
  {B{\"o}hringer} \& {Collins}}{{Ortiz-Gil} et~al.}{2004}]{ort04}
{Ortiz-Gil} A.,  {Guzzo} L.,  {Schuecker} P.,  {B{\"o}hringer} H.,    {Collins}
  C.~A.,  2004, MNRAS, 348, 325

\bibitem[\protect\citeauthoryear{{Peres}, {Fabian}, {Edge}, {Allen},
  {Johnstone} \& {White}}{{Peres} et~al.}{1998}]{per98}
{Peres} C.~B.,  {Fabian} A.~C.,  {Edge} A.~C.,  {Allen} S.~W.,  {Johnstone}
  R.~M.,    {White} D.~A.,  1998, MNRAS, 298, 416

\bibitem[\protect\citeauthoryear{{Peterson et~al.}}{{Peterson
  et~al.}}{2001}]{pet01}
{Peterson J.~R. et~al.} 2001, A\&A, 365, L104

\bibitem[\protect\citeauthoryear{{Ponman}, {Cannon} \& {Navarro}}{{Ponman}
  et~al.}{1999}]{pon99}
{Ponman} T.~J.,  {Cannon} D.~B.,    {Navarro} J.~F.,  1999, Nat, 397, 135

\bibitem[\protect\citeauthoryear{{Pope}, {Pavlovski}, {Kaiser} \&
  {Fangohr}}{{Pope} et~al.}{2006}]{pop06a}
{Pope} E.~C.~D.,  {Pavlovski} G.,  {Kaiser} C.~R.,    {Fangohr} H.,  2006,
  MNRAS, 367, 1121

\bibitem[\protect\citeauthoryear{{Popesso}, {Biviano}, {B{\"o}hringer},
  {Romaniello} \& {Voges}}{{Popesso} et~al.}{2005}]{pop05}
{Popesso} P.,  {Biviano} A.,  {B{\"o}hringer} H.,  {Romaniello} M.,    {Voges}
  W.,  2005, A\&A, 433, 431

\bibitem[\protect\citeauthoryear{Prestage \& Peacock}{Prestage \&
  Peacock}{1988}]{pre88}
Prestage R.~M.,  Peacock J.~A.,  1988, MNRAS, 230, 131

\bibitem[\protect\citeauthoryear{{Quilis}, {Bower} \& {Balogh}}{{Quilis}
  et~al.}{2001}]{qui01}
{Quilis} V.,  {Bower} R.~G.,    {Balogh} M.~L.,  2001, MNRAS, 328, 1091

\bibitem[\protect\citeauthoryear{{Rafferty}, {McNamara}, {Wise} \&
  {Nulsen}}{{Rafferty} et~al.}{2006}]{raf06}
{Rafferty} D.~A.,  {McNamara} B.~R.,  {Wise} M.~W.,    {Nulsen} P.~E.~J.,
  2006, ApJ, submitted: astro-ph/0605323

\bibitem[\protect\citeauthoryear{{Reynolds}, {McKernan}, {Fabian}, {Stone} \&
  {Vernaleo}}{{Reynolds} et~al.}{2005}]{rey05}
{Reynolds} C.~S.,  {McKernan} B.,  {Fabian} A.~C.,  {Stone} J.~M.,
  {Vernaleo} J.~C.,  2005, MNRAS, 357, 242

\bibitem[\protect\citeauthoryear{{Roychowdhury}, {Ruszkowski} \&
  {Nath}}{{Roychowdhury} et~al.}{2005}]{roy05}
{Roychowdhury} S.,  {Ruszkowski} M.,    {Nath} B.~B.,  2005, ApJ, 634, 90

\bibitem[\protect\citeauthoryear{{Ruszkowski} \& {Begelman}}{{Ruszkowski} \&
  {Begelman}}{2002}]{rus02}
{Ruszkowski} M.,  {Begelman} M.~C.,  2002, ApJ, 581, 223

\bibitem[\protect\citeauthoryear{{Ruszkowski}, {Br{\"u}ggen} \&
  {Begelman}}{{Ruszkowski} et~al.}{2004}]{rus04}
{Ruszkowski} M.,  {Br{\"u}ggen} M.,    {Begelman} M.~C.,  2004, ApJ, 611, 158

\bibitem[\protect\citeauthoryear{Scannapieco, Silk \& Bouwens}{Scannapieco
  et~al.}{2005}]{sca05}
Scannapieco E.,  Silk J.,    Bouwens R.,  2005, ApJ, 365, L13

\bibitem[\protect\citeauthoryear{Schombert}{Schombert}{1986}]{sch86}
Schombert J.~M.,  1986, ApJ Supp., 60, 603

\bibitem[\protect\citeauthoryear{{Spitzer}}{{Spitzer}}{1962}]{spi62}
{Spitzer} L.,  1962, {Physics of Fully Ionized Gases}.
Wiley, New York

\bibitem[\protect\citeauthoryear{{Stoughton et~al.}}{{Stoughton
  et~al.}}{2002}]{sto02}
{Stoughton C. et~al.} 2002, AJ, 123, 485

\bibitem[\protect\citeauthoryear{{Strauss et~al.}}{{Strauss
  et~al.}}{2002}]{str02}
{Strauss M.~A. et~al.} 2002, AJ, 124, 1810

\bibitem[\protect\citeauthoryear{{Sun}, {Forman}, {Vikhlinin}, {Hornstrup},
  {Jones} \& {Murray}}{{Sun} et~al.}{2003}]{sun03}
{Sun} M.,  {Forman} W.,  {Vikhlinin} A.,  {Hornstrup} A.,  {Jones} C.,
  {Murray} S.~S.,  2003, ApJ, 598, 250

\bibitem[\protect\citeauthoryear{{Tamura et~al.}}{{Tamura
  et~al.}}{2001}]{tam01}
{Tamura T. et~al.} 2001, A\&A, 365, L87

\bibitem[\protect\citeauthoryear{{Tremaine et~al.}}{{Tremaine
  et~al.}}{2002}]{tre02}
{Tremaine S. et~al.} 2002, ApJ, 574, 740

\bibitem[\protect\citeauthoryear{{Valentijn} \& {Bijleveld}}{{Valentijn} \&
  {Bijleveld}}{1983}]{val83}
{Valentijn} E.~A.,  {Bijleveld} W.,  1983, A\&A, 125, 223

\bibitem[\protect\citeauthoryear{{Voigt} \& {Fabian}}{{Voigt} \&
  {Fabian}}{2004}]{voi04}
{Voigt} L.~M.,  {Fabian} A.~C.,  2004, MNRAS, 347, 1130

\bibitem[\protect\citeauthoryear{{Voigt}, {Schmidt}, {Fabian}, {Allen} \&
  {Johnstone}}{{Voigt} et~al.}{2002}]{voi02}
{Voigt} L.~M.,  {Schmidt} R.~W.,  {Fabian} A.~C.,  {Allen} S.~W.,
  {Johnstone} R.~M.,  2002, MNRAS, 335, L7

\bibitem[\protect\citeauthoryear{{Voit} \& {Donahue}}{{Voit} \&
  {Donahue}}{1997}]{voi97}
{Voit} G.~M.,  {Donahue} M.,  1997, ApJ, 486, 242

\bibitem[\protect\citeauthoryear{{von der Linden}, Best, Kauffmann \&
  White}{{von der Linden} et~al.}{2006}]{lin06}
{von der Linden} A.,  Best P.~N.,  Kauffmann G.,    White S. D.~M.,  2006,
  MNRAS, submitted; astro-ph/0611196

\bibitem[\protect\citeauthoryear{{Wu}, {Fabian} \& {Nulsen}}{{Wu}
  et~al.}{2000}]{wu00}
{Wu} K.~K.~S.,  {Fabian} A.~C.,    {Nulsen} P.~E.~J.,  2000, MNRAS, 318, 889

\bibitem[\protect\citeauthoryear{{Xue}, {B{\"o}hringer} \& {Matsushita}}{{Xue}
  et~al.}{2004}]{xue04}
{Xue} Y.-J.,  {B{\"o}hringer} H.,    {Matsushita} K.,  2004, A\&A, 420, 833

\bibitem[\protect\citeauthoryear{{Xue} \& {Wu}}{{Xue} \& {Wu}}{2000}]{xue00}
{Xue} Y.-J.,  {Wu} X.-P.,  2000, ApJ, 538, 65

\bibitem[\protect\citeauthoryear{{York et~al.}}{{York et~al.}}{2000}]{yor00}
{York D.~G. et~al.} 2000, AJ, 120, 1579

\bibitem[\protect\citeauthoryear{{Zakamska} \& {Narayan}}{{Zakamska} \&
  {Narayan}}{2003}]{zak03}
{Zakamska} N.~L.,  {Narayan} R.,  2003, ApJ, 582, 162

\end{thebibliography}
\bibliographystyle{mn2e}

\begin{appendix}
\section{The heating rate due to thermal conductivity}
\label{apptherm}

The volume heating rate due to thermal conductivity from an
approximately infinite heat bath outside a radius $r$ is given by

\begin{equation}
\label{eqtherm}
\epsilon = \frac{1}{r^2} \frac{\partial}{\partial r} \left( r^2 \kappa
\frac{\partial T}{\partial r} \right). 
\end{equation}

\noindent where $\kappa$ is the thermal conductivity of the gas and $T$ is
its temperature. \citet{spi62} calculated the thermal conductivity for a
pure hydrogen gas to be:

\begin{equation}
\label{eqkappa}
\kappa _{\rm S} = 1.84 \times 10^{-10} \frac{T^{5/2}}{\ln \Lambda}\,{\rm
W\,m^{-1}\,K^{-1}}
\end{equation}

\noindent where $\ln \Lambda$ is the Coulomb logarithm. In the presence of
magnetic fields, however, the thermal conductivity will be lower. This is
usually accounted for by incorporating a factor $f_{\rm Spitz}$, such that

\begin{equation}
\kappa = \kappa _{\rm S} f_{\rm Spitz}.
\end{equation}

The Coulomb logarithm in Equation~\ref{eqkappa} depends weakly on the gas
density and temperature, but to a good approximation can be treated as a
constant with $\ln \Lambda \approx 37$ for typical conditions in clusters.
Hence, 

\begin{equation}
\label{eqkappa2}
\kappa = \kappa_0 T^{5/2},
\end{equation}

\noindent where $\kappa_0 \approx 5\times 10^{-12} f_{\rm Spitz}$.

\citet{all01} showed that the Chandra data of massive relaxed clusters
could be fitted by a `universal' temperature profile:

\begin{equation}
\label{eqtemp}
T = T_{2500} \left[ T_0 + T_1 \frac{\left( x/ x_{\rm c}
\right)^{\eta}}{1+\left( x/ x_{\rm c} \right)^{\eta}} \right], 
\end{equation}

\noindent where $T_0=0.40$, $T_1=0.61$, $x_{\rm c}=0.087$ and $\eta=1.9$
are dimensionless constants that they fitted to their data, $T_{2500}$ is
the temperature of the cluster gas at the radius $r_{2500}$ (within which
the mass density is 2500 times the critical density of the universe), and
the dimensionless parameter $x$ is given by $x=r /
r_{2500}$. \citet{all01} assume $r_{2500} \approx 0.3 r_{200}$.

Substituting Equations~\ref{eqkappa2} and~\ref{eqtemp} for $\kappa$ and
$T$ into Equation~\ref{eqtherm} gives:

\begin{eqnarray}
\label{eqeps}
\epsilon & = & \Bigg\{ \eta T_1 T_{2500}^{7/2}  \left( \frac{x}{x_{\rm c}}
 \right)^{\eta} \kappa _0 \left[ T_0 +T_1 \left( 1-
 \frac{1}{1+\left( x / x_{\rm c} 0\right)^{\eta}} \right)\right]^{0.5}
  \nonumber \\ 
 & & \left[ T_0 + \left( T_0 + T_1 \right) \left( \frac{x}{x_{\rm c}}
 \right)^{\eta} \right] \bigg[ 2 \left( 1 - \eta \right) \left( T_0 + T_1 \right) \left(
 \frac{x}{x_{\rm c}} \right)^{2 \eta} \nonumber\\
 & & \left. + \left( 4 T_0 + \left( 2 + 7 \eta \right) T_1 \right) \left( \frac{x}{x_{\rm c}}
 \right)^{\eta} + 2 \left(1+ \eta \right) T_0 \bigg]^{}\Bigg\} \right/ \nonumber\\ 
 & &  \left\{ 2 r^2 \left[ 1 + \left( \frac{x}{x_{\rm c}} \right)^{\eta}
 \right]^5 \right\}. 
\end{eqnarray}

The total heating power due to thermal conduction inside a radius $r$ is
then roughly $4 \pi r^3 \epsilon(r) / 3$. Setting $r = r_{\rm cool}$, and
using $r_{\rm cool} \approx 67 (\sigma_{\rm v} / 500$\,km\,s$^{-1})$\,kpc
from Section~\ref{bcgimport} then gives an approximation for the flow of
heat to within the cooling radius.

In order to calculate this, $r_{\rm 200}$ and $T_{\rm 2500}$ are
required. $r_{\rm 200}$ can be estimated from the velocity dispersion as
$r_{\rm 200} \approx 1230 (\sigma_{\rm v} / 500$\,km\,s$^{-1})$\,kpc
\citep[from][appropriate for redshift zero and the presently adopted
cosmology]{fin05}. This then gives

\begin{equation}
 x = 0.18 \left( \frac{\sigma_{\rm v}}{500\,{\rm km\,s^{-1}}} \right)^{-0.55}.
\end{equation}

Similarly, $T_{2500}$ depends on the velocity dispersion, and can be
estimated from the $T-\sigma_{\rm v}$ relation for clusters \citep{xue00}:

\begin{equation}
 T_{\rm 2500} \approx 2.4 \times 10^7 \left( \frac{\sigma_{\rm v}}{500\,{\rm
 km\,s^{-1}}} \right)^{1.56} K.
\end{equation}

Substituting for $x$ and $T_{2500}$ into Equation~\ref{eqeps} allows the
total heating rate within the cooling radius due to thermal conduction to
be estimated, for different values of $f_{\rm Spitz}$. Note that strictly
these are only valid for systems with velocity dispersions $\sigma_{\rm v}
\gta 800$km\,s$^{-1}$, since the `universal' temperature profile of
\citet{all01} was only determined from massive clusters and may not fit
lower mass systems. Nevertheless, it provides at least an indication of
the likely importance of thermal conduction within systems of different
masses, as shown on Figure~\ref{coolfrac}.

\end{appendix}

\label{lastpage}
\end{document}